\documentclass[preprint,authoryear,5p,times,twocolumn]{elsarticle}
\usepackage{amsmath}
\usepackage{graphicx}
\usepackage{amssymb}
\usepackage{upgreek}
\usepackage{bbm}
\usepackage{dsfont}
\usepackage{sansmath}
\def\d{{\rm d}}
\def\D{{\rm D}}
\def\del{\partial}
\def\grad{\mbox{\boldmath{$\nabla$}}}
\def\bfnabla{\mbox{\boldmath{$\nabla$}}}
\def\grad-s{\mbox{\boldmath{$\nabla$}}_{\!\! s}\,}
\def\bfx{{\bf x}}

\def\bfv{{\bf v}}
\def\bfk{{\bf k}}
\def\bfOmega{\mbox{\boldmath{$\Omega$}}}

\def\bfCalF{\mbox{\boldmath{${\cal F}$}}}
\def\bfCalD{\mbox{\boldmath{${\cal D}$}}}
\def\sub#1{\mbox{\tiny $\mbox{\sc {#1}}$}}

\newcommand{\degr}{\ensuremath{^\circ}}

\usepackage{color}
\definecolor{com_red}{rgb}{.8,.2,0.1}
\definecolor{ins_green}{rgb}{.1,.5,0.1}
\definecolor{ins_red}{rgb}{.8,.1,0.1}

\journal{Icarus}
\begin{document}
\begin{frontmatter}
\title{Intercomparison of General Circulation Models for
  Hot Extrasolar Planets}

\author[label2]{I.~Polichtchouk\corref{cor1}}
\ead{i.polichtchouk@qmul.ac.uk}
\author[label2]{J.~Y-K.~Cho }
\author[label2]{C.~Watkins }
\author[label2,label3]{H.~Th.~Thrastarson }
\author[label2,label4]{O.~M.~Umurhan}
\author[label3]{M.~de~la~Torre~Ju{\'a}rez }
\cortext[cor1]{Corresponding author}
\address[label2]{School of Physics and Astronomy, Queen Mary 
University of London, London E1 4NS, UK}
\address[label3]{Jet Propulsion Laboratory, California Institute of 
Technology, Pasadena, CA 91109, USA}
\address[label4]{School of Natural Sciences, University of California, 
Merced, CA 95343, USA}

\begin{abstract}
  We compare five general circulation models (GCMs) which have been
  recently used to study hot extrasolar planet atmospheres (BOB, CAM,
  IGCM, MITgcm, and PEQMOD), under three test cases useful for
  assessing model convergence and accuracy.  Such a broad, detailed
  intercomparison has not been performed thus far for extrasolar
  planets study.  The models considered all solve the traditional
  primitive equations, but employ different numerical algorithms or
  grids (e.g., pseudospectral and finite volume, with the latter
  separately in longitude-latitude and `cubed-sphere' grids).  The
  test cases are chosen to cleanly address specific aspects of the
  behaviors typically reported in hot extrasolar planet simulations:
  {\it 1})~steady-state, {\it 2})~nonlinearly evolving baroclinic
  wave, and {\it 3})~response to fast timescale thermal relaxation.
  When initialized with a steady jet, all models maintain the
  steadiness, as they should---except MITgcm in cubed-sphere grid.  A
  very good agreement is obtained for a baroclinic wave evolving from
  an initial instability in pseudospectral models (only).  However,
  exact numerical convergence is still not achieved across the
  pseudospectral models: amplitudes and phases are observably different.
  When subject to a typical `hot-Jupiter'-like forcing, all five
  models show quantitatively different behavior---although
  qualitatively similar, time-variable, quadrupole-dominated flows are
  produced.  Hence, as have been advocated in several past studies,
  specific quantitative predictions (such as the location of large
  vortices and hot regions) by GCMs should be viewed with caution.
  Overall, in the tests considered here, pseudospectral models in pressure
  coordinate (PEBOB and PEQMOD) perform the best and MITgcm in
  cubed-sphere grid performs the worst.
\end{abstract}
\label{firstpage}
\end{frontmatter}
\section{Introduction}\label{intro}

Carefully testing general circulation models (GCMs) of extrasolar
planets is important for understanding the physical properties of the
atmospheres and for attaining confidence in the complex models
themselves.  Intercomparison of full GCMs, as well as benchmarking of
dynamical cores and testbed models against `standard solutions', are
common in Earth studies
\citep[e.g.,][]{heldSuarez,BoerDenis97,Polvani04,Jablonowski06}.
Intercomparisons are also becoming more common for circulation models
of other Solar System planets \citep[e.g.,][]{Lebonnois13}.  However,
similar intercomparisons have not been performed for models of hot
extrasolar planets.  Given that the conditions of many extrasolar
planets are markedly different than the Earth---and much more exacting
on the circulation models---it is useful to subject the models to
tests which are appropriate for extrasolar conditions
\citep[e.g.,][]{Thrastarson11}.

Thus far, only \citet{Rauscher10} and \citet{Heng11} have explicitly
attempted to intercompare simulations of hot extrasolar planets
performed with different GCMs.  The former study attempts to compare
their results using the Intermediate General Circulation Model
\citep{Blackburn85} with those reported in \citet{Cooper05} using the
ARIES/GEOS model \citep{Suarez95}.  However, while qualitatively
similar features were observed, the comparison was somewhat
inconclusive because the model setup was not identical.  In their
studies using the Community Atmosphere Model \citep{Collins04},
\citet{Thrastarson10,Thrastarson11} have shown sensitivity to initial
condition, as well as thermal relaxation and explicit numerical
dissipation specifications.  A clearer comparison than in
\citet{Rauscher10} has been presented in \citet{Heng11}.  In the
latter study, time-mean zonally-averaged (i.e.,
longitudinally-averaged) fields are presented from simulations with
the Flexible Modeling System \citep{Anderson04}, using two different
types of numerical algorithm (pseudospectral and finite volume).
However, while zonal and temporal averaging is somewhat justifiable
for rapidly rotating planets, the procedure is less useful for the
more slowly rotating planets, such as those considered in the study:
the averaging can destroy dynamically-significant
flow structures, as well as conceal subtle numerical and coding
errors. \footnote{During the
      preparation of this manuscript another study, by
      \cite{Bending13} appeared that compares their results with those
      of \cite{Menou09}. The authors of the new study report that they
      are not able to reproduce precisely the results of the older
      study, although both studies use the same dynamical core
      (Section~\ref{cores}).}

In addition to the setup not being same or systematic across different
models, the inconclusiveness of the past comparisons and the general
variability of the model results stem from the fact that the models
employ different numerical algorithms, grids, and coordinates to solve
the governing equations---as we shall show in this work.  Moreover,
the numerical parameters of the model calculations are often not
described explicitly in the literature, or even in the technical
documentations of the models themselves.  Thus, it has been difficult
to ascertain which differences between model outputs are due to the
model and which are due to the setup.  Here, we perform a careful
comparison of five GCMs recently used to study hot extrasolar planet
atmospheres.  The GCMs are: BOB\footnote{{\it Built on Beowolf}
  \ \ \citep{Scott03}}, CAM\footnote{{\it Community Atmosphere Model
    -- version 3.0} \ \ \citep{Collins04}}, IGCM\footnote{{\it
    Intermediate General Circulation Model} \ \ \citep{Blackburn85}},
MITgcm\footnote{{\it MIT general circulation model -- checkpoint64d}
  \ \ \citep{Adcroft12}}, and PEQMOD\footnote{{\it Primitive Equations
    Model} \ \ \citep{Saravanan92}}.  They have been used, for
example, in the following extrasolar planet circulation studies:
BOB~\citep{Beaulieu11,Polichtchouk12}, CAM
\citep{Thrastarson10,Thrastarson11}, IGCM \citep{Menou09,Rauscher10},
MITgcm \citep{Showman09,Lewis10}, and PEQMOD (Cho and Polichtchouk, in
prep.).

The five GCMs are submitted to three tests which are useful for
assessing model convergence and accuracy.  The tests are chosen to
specifically address three features that have been typically reported
in hot extrasolar planet atmospheric flow simulations: {\it 1)}~steady
flow, {\it 2)}~nonlinear baroclinic wave, and {\it 3)}~response to a
fast timescale thermal relaxation.  We stress that, in addition to
their good range and relevance, the tests are purposely chosen with
reproducibility of the results in mind: the tests are not difficult to
set up and full descriptions of the test cases (as well as the GCMs
tested) are provided, along with all of the model parameter values
used in the simulations (see Appendix)---as per our usual practice.
We are also happy to share all source codes and input files/parameters
used in this study.  Note that the emphasis in this work is on models
tested in their `default configuration' (i.e., essentially as they are
unpacked), modulo minor modifications to facilitate equatable (as well
as equitable)\footnote{Equitable refers to `impartial' or `fair', and
  equatable refers to `equivalent' or `comparable'.}  comparisons.

The overall plan of the paper is as follows.  In
Section~\ref{cores_and_tests}, we review the governing equations
solved by the five GCMs and describe the discretization and
dissipation schemes used in the models.  In Section~\ref{results}, the
three test cases are carefully described and the results from the
tests are presented in turn.  Both {\it inter}-model and {\it
  intra}-model comparisons are presented in detail, where the former
comparison refers to `between different models' and the latter
comparison refers to `within a single model'.  The aim of this
section---indeed, of this entire paper---is to permit one to go beyond
broad-brush comparisons based on strongly dissipated/constrained or
averaged fields.  In Section~\ref{summary}, summary and conclusions
are given, along with some discussion of implications of this work.

\section{Dynamical Cores and Test Cases}\label{cores_and_tests}

\subsection{Dynamical Cores}\label{cores}

The GCMs---or, more precisely, their `dynamical cores'---discussed in
this work all solve the hydrostatic primitive equations for the `dry'
atmosphere.  The dynamical core is essentially that part of the GCM
which remains when all the sophisticated physical parameterizations
(e.g., convection, radiation, wave-drag, etc.) have been stripped
away: it is the engine of the GCM.  In this paper, we refer to `GCMs'
and `dynamical cores' interchangeably, as the distinction is not
particularly important here.  None of the sophisticated physical
parameterizations are used in any of the models for the comparisons:
only a crude heating/cooling scheme is used in one of the test cases.
In general, it is prudent to test and characterize the core before
moving onto the full GCM.\footnote{Note that, in comparisons of full
  GCMs for the Earth, model differences generally increase when
  physics parameterizations are included (e.g., \citet{Blackburn13})}.

The equations solved govern the large-scale dynamics of planetary
atmospheres (e.g., \citet{Holton92}; see also \citet{Cho08b} for some
discussions relevant to the current work).  Given that the GCMs tested
solve the equations in different vertical coordinate systems (e.g.,
pressure, sigma, eta---see below), we first present and discuss the
equations in the generalized vertical coordinate, $s$.  In the
$s$-coordinate, the hydrostatic primitive equations read
\begin{subequations}\label{PE general}
 \begin{eqnarray}
   \lefteqn{ \frac{\D \bfv}{\D t}\  =\ 
     -\frac{1}{\rho}\grad-s p - g\grad-s z - f\bfk\!\times\!\bfv +
     \bfCalF_{\!\mbox{\tiny $\bfv$}} + \bfCalD_{\mbox{\tiny $\bfv$}}}\\
   \lefteqn{ \,\frac{\D \theta}{\D t}\ =\ 
     \frac{\theta}{c_p T}\,\dot{q}_{\rm net} + 
     {\cal D_{\mbox{\tiny $\theta$}}} }\\
   \lefteqn{\  \frac{\del p}{\del s}\ =\ 
     -\rho g\,\frac{\del z}{\del s} }\\
   \lefteqn{\ \ \  0\ \, \ = \ \frac{\del}{\del s}\!\!
     \left(\frac{\del p}{\del t}\right)_{\!\! s} + 
     \grad-s\!\! \cdot\! \left(\bfv\frac{\del p}{\del s}\right) +
     \frac{\del }{\del s}\!\!\left(\dot{s}
       \frac{\del p}{\del s}\right)\, , }\\
   \lefteqn{ \mbox{where} \nonumber }\\
   \lefteqn{ \qquad \qquad \frac{\D\ \ }{\D t}\ \equiv\  
     \left(\frac{\del}{\del t} \right)_{\! s} + \bfv\! \cdot\! 
     \grad-s + \dot{s} \frac{\partial}{\partial s}\ \ \ . \nonumber }
 \end{eqnarray}
\end{subequations}
Here, $\bfv(\bfx,s,t) = (u, v)$ is the (zonal, meridional) velocity in
the frame rotating with $\bfOmega$, the planetary rotation vector, and
${\bf x}\in\mathbb{R}^2$; $\dot{s} \equiv \D s/\D t$ is the
generalized vertical velocity; $z = z(\bfx,s,t)$ is the physical
height, directed locally upward (in the direction of the unit vector
${\bf k}$); $\grad-s$ is the two-dimensional (2D) gradient operator,
operating along constant surfaces of $s = s(\bfx,z,t)$;
$\rho(\bfx,s,t)$ is the density; $p(\bfx,s,t)$ is the pressure; $f = 2
\Omega \sin\phi = 2\bfOmega\!\cdot\!\bfk$ is the Coriolis parameter,
where $\phi$ is the latitude;
$\bfCalF_{\!\mbox{\tiny$\bfv$}}(\bfx,s,t)$ represents momentum
sources; $\bfCalD_{\!\mbox{\tiny$\bfv$}}(\bfx,s,t)$ and ${\cal
  D}_{\mbox{\tiny$\theta$}}(\bfx,s,t)$ represent momentum and
potential temperature sinks, respectively; $g$ is the gravitational
acceleration, assumed to be constant and to include the centrifugal
acceleration contribution; $\theta(\bfx,s,t) = T(p_{\rm
  r}/p)^{\kappa}$ is the potential temperature, where $T(\bfx,s,t)$ is
the temperature, $p_{\rm r}$ is a constant reference pressure, and
$\kappa = {\cal R}/c_p$, with $\cal R$ the specific gas constant and
$c_p$ the constant specific heat at constant pressure; and,
$\dot{q}_{\rm net}(\bfx,s,t)$ is the {\it net} diabatic heating rate
(i.e., heating minus cooling).

The set of equations, (\ref{PE general}a)--(\ref{PE general}d), is
closed with the ideal gas equation of state, $p = \rho {\cal R} T$.
The equation set is also supplemented with the boundary conditions,
\begin{subequations}\label{BC}
\begin{eqnarray}
  \lefteqn{ \ \ \dot{s}\ =\ 0 \hspace*{3cm} {\rm at} \qquad  
    s = s_{\sub{T}} }\\
  \lefteqn{ \ \ \dot{s}\ =\ \frac{\del s_{\sub{B}}}{\del t} +
    {\bf v}_{\sub{B}}\!\cdot\!\bfnabla s_{\sub{B}} \hspace*{1.25cm} 
    {\rm at} \qquad s = s_{\sub{B}}\, . }
\end{eqnarray}
\end{subequations}
Here, $s_{\sub{T}}$ is the boundary surface at the top; $s_{\sub{B}}$
is the boundary surface at the bottom, at a fixed altitude above the
reference height ($z = 0$); and, ${\bf v}_{\sub{B}}$ is horizontal
velocity at the bottom surface.  Boundary conditions~(\ref{BC}) imply
no mass transport through the upper and lower boundary surfaces.
Note, if the lower boundary coincides with a constant $s$-surface
(i.e., $s_{\sub{B}} \ne s_{\sub{B}}(\bfx,t)$), then the boundary
condition~(\ref{BC}b) simply reduces to
\begin{eqnarray}
 \lefteqn{ \ \ \dot{s}\ =\ 0 \hspace*{3cm} {\rm at} \qquad  
   s = s_{\sub{B}}\, . }
\end{eqnarray}

While letting $s \rightarrow z$ might be an intuitive choice for a
vertical coordinate, it is common in GCMs to use pressure or other,
pressure-based, coordinates: for example, $s \rightarrow p$, $s
\rightarrow \sigma(p)$, or $s \rightarrow \eta(p)$.  In the
$p$-coordinate, the continuity equation (\ref{PE general}d) takes on a
simple diagnostic form.  However, the coordinate system poses a
computational disadvantage when modelling a planet with a solid
surface, if topography is present.  In this case, the boundary
condition (\ref{BC}b) becomes difficult to handle.  The
$\sigma$-coordinate or the $\eta$-coordinate circumvents this problem
because the planet's surface does not intersect a vertical coordinate
surface in either coordinate systems.

All the cores tested solve the equations in the Eulerian framework.
Hence, for all of them, there is an associated grid for the
computational domain---e.g., longitude-latitude (LL) and
cubed-sphere~(CS) grids for the MITgcm core and the Gaussian grid for
the remaining four cores.  The numerical integrations are directly
performed on the LL and CS grids in MITgcm, while only the nonlinear
products are evaluated and initial conditions are specified on the
Gaussian grid in the pseudospectral cores.  More details on the cores
are provided below, beginning with the pseudospectral ones.  All cores
use common values for the planetary parameters, which are listed in
Table~1: the values characteristic of the planet HD209458b are used.

\subsubsection{Pseudospectral Cores}\label{PS cores}

BOB, CAM, IGCM, and PEQMOD cores use the highly-accurate
pseudospectral algorithm \citep{Orszag70,Eliasen70,Canuto88} for the
horizontal direction.  Equations~(\ref{PE general}) in the
vorticity-divergence form are transformed with triangular truncation
(i.e., $M = N$, where $M$ is the maximum zonal wavenumber and $N$ is
the maximum total wavenumber retained in the spherical harmonic
expansion).\footnote{IGCM uses the `jagged triangular truncation', in
  which $n \leq M-1$ for variables even about the equator and $n \leq
  M$ for variables odd about the equator.  Here, $n$ is the total
  wavenumber.}  Given $M$ and $N$, all the nonlinear products in the
full set of equations~(\ref{PE general}) are first evaluated in
physical space on a Gaussian grid with enough points, in principle, to
eliminate aliasing errors and then transformed to spectral space.  The
linear terms are directly transformed.

For the vertical direction, a standard second-order finite difference
scheme is used.  In this direction, the grid is typically equally
spaced in $p$-coordinates or $\sigma$-coordinates, as are all the
simulations in this work (see below).  Note that all the cores tested
in this work use the Lorenz grid \citep{Lorenz60}, in which the
vertical velocity is defined at the boundary of the layers and the
prognostic variables (e.g., vorticity, divergence, and
temperature/potential temperature) are defined at the centers of the
layers.  The grid allows boundary conditions of no flux at the top and
bottom of the domain to be easily fulfilled.  However, a spurious
computational mode is admitted, arising from an extra degree of
freedom introduced in the potential temperature \citep{Arakawa88}.  We
have observed that this can lead to small-amplitude oscillations in
the temperature field on the timescale of a timestep.
\begin{table}
  \centering
  \caption{Parameter values for the hot extrasolar planet HD209458b.
    Here, $g$ is surface gravity; $R_p$ is equatorial radius; $\Omega$
    is rotation rate; ${\cal R}$ is gas constant; $c_p$ is specific
    heat at constant pressure; $H$ is scale height; $T_{\rm eq}$ is
    equilibrium temperature; and, $N$ is Brunt-V\"{a}is\"{a}l\"{a}
    frequency..}
    \label{table-params}
    \vspace*{.2cm}
    \begin{tabular}{lll} 
      \\
      \hline
      Parameter \hspace*{3mm} & Value \hspace*{1cm} & Units \\
      \hline
      $g$   & 9.8  &  m~s$^{-2}$ \\
      $R_{p}$   &  $10^{8}$     & m \\
      $\Omega$   & $2.1\!\times\! 10^{-5}$    &  s$^{-1}$ \\
      ${\cal R}$  & $3.5\!\times\! 10^3$     & J~kg$^{-1}$~K$^{-1}$  \\
      $c_{p}$   & $1.23\!\times\! 10^4$     &   J~kg$^{-1}$K$^{-1}$   \\
      $H$   & $1.23\!\times\! 10^4$     &   J~kg$^{-1}$K$^{-1}$   \\
      $T_{\rm eq}$   & $1.23\!\times\! 10^4$     &   J~kg$^{-1}$K$^{-1}$   \\
      $N$   & $1.23\!\times\! 10^4$     &   J~kg$^{-1}$K$^{-1}$   \\
      \hline\\
    \end{tabular}
\end{table}

As for the time integration, the above four cores use a semi-implicit
leap-frog scheme.  A small Robert-Asselin time filter coefficient
$\epsilon$ \citep{Robert66, Asselin} is applied at every timestep in
each layer to filter the computational mode arising from using the
second-order time-marching scheme \citep[see, e.g.,][]{Thrastarson10}.
To integrate the equations over long simulation durations, explicit
dissipation is applied to the prognostic variables so that artificial
accumulation of energy at small scales is prevented
\citep[e.g.,][]{Cho96, Thrastarson11}.  The dissipations, ${\cal
  D}_{{\bf v}, \theta}$ in equations~(\ref{PE general}), are in the
form of a `hyperdissipation' operator (see equation~(\ref{hyper})
below).
\\
BOB and PEQMOD solve equations~(\ref{PE general}) in the
$p$-coordinate:
\begin{subequations}\label{PE bob}
 \begin{eqnarray}
   \lefteqn{ \frac{\partial{\zeta}}{\partial t}\ \, =\ 
     \bfk \cdot \bfnabla\!\times\! {\bf n_{\rm p}} + \cal{D}_{\zeta} }\\
   \lefteqn{ \frac{\partial{\delta}}{\partial t}\ \, = \ 
     \bfnabla \cdot {\bf n_{\rm p}} - \nabla^2\big(E + \Phi\big) + 
     \cal{D}_{\delta}}\\
   \lefteqn{ \frac{\partial \Phi}{\partial \xi}\, =\ -c_p\, \theta}\\
   \lefteqn{ \frac{\partial \omega}{\partial p}\ = - \delta}\\
   \lefteqn{ \frac{\partial{\theta}}{\partial t}\ =\   
     -\bfnabla\! \cdot\! \big(\theta\, \bfv\big) - 
     \frac{\del (\omega\theta)}{\del p} + 
     \frac{\theta}{c_p T}\dot{q}_{\rm net}+\cal{D}_{\theta}\, , }
 \end{eqnarray}
\end{subequations}   
where ${\bf \zeta} = \bfk\cdot(\bfnabla\times\bfv)$ is the relative
vorticity;
\[ {\bf n_{\rm p}}\ =\ -\big( \zeta + f \big)\bfk\times\bfv -
\delta\bfv - \frac{\del(\omega \bfv)}{\del p}\, ;
\]
$\delta = \bfnabla\cdot\bfv$ is the divergence; $E = ({\bf v} \cdot
{\bf v})/2$ is the specific kinetic energy; $\Phi = gz$ is the
specific geopotential above the planetary radius $R_p$; $\omega = \D p
/ \D t$ is the vertical velocity, where
\[\frac{\D}{\D t} \equiv \frac{\del}{\del t} + \bfv\!
\cdot\! \bfnabla + \omega\frac{\del}{\del p}
\] is the material derivative with $\bfnabla$ operating along constant
surfaces of $p$; and, the diffusion terms $\cal{D}_{\chi}$, where
$\chi = \{\zeta,\delta, \theta\}$, are given by
\citep[e.g.,][]{Cho96}:
\begin{equation}\label{hyper} {\cal D}_{\chi}\ =\ \nu_{2
    \mathfrak{p}}\big[(-1)^{\mathfrak{p}+1}\nabla^{2\mathfrak{p}} +
  {\cal C}\big]\ \chi\, ,
\end{equation}
where $\mathfrak{p}$ (different from $p$, the pressure) is the order
of the hyperdissipation operator; $\nu_{2\mathfrak{p}}$ is the
hyperdissipation coefficient; and, ${\cal C} =
(2/R_p^2)^{\,\mathfrak{p}}$ is a correction term added to the
vorticity and divergence equations to prevent damping of uniform
rotations for angular momentum conservation.  Note that, in the
baroclinic wave test case, $\mathfrak{p} = 1$, leading to the normal
Laplacian dissipation.  In the diabatic test case, $\mathfrak{p} = 2$,
leading to the more scale-selective `superdissipation'.

BOB and PEQMOD have an additional constraint of no vertically
integrated divergence over the whole atmosphere.  This constraint
excludes the divergent `shallow-water mode', which has a barotropic
vertical structure, and increases the computationally stability of the
models.  With this additional constraint, the boundary
conditions~(\ref{BC}) become
\begin{equation}\label{BC bob}
\omega = 0 \qquad {\rm{at}} \qquad p = 0, \ p_{\sub{B}},
\end{equation}
and the lower boundary always coincides with a constant $p$-surface.
The above boundary conditions entail zero flux of any quantity through
the upper and lower pressure surfaces.

As already discussed, in the $p$-coordinate continuity
equation~(\ref{PE general}c) simplifies to a simple diagnostic
equation~(\ref{PE bob}d).  Thus, with the boundary conditions~(\ref{BC
  bob}), BOB and PEQMOD actually only integrate three equations---that
is, only three variables are prognostic.  The vertical discretization
scheme in BOB and PEQMOD preserves the global conservation properties
of absolute angular momentum, potential temperature and total energy
in the absence of forcing/dissipation.  Note, unlike BOB, PEQMOD
implements a slightly non-standard Gaussian transform grid, for which
the equatorial point is constrained to be one of the Gaussian points.

IGCM is formulated in the $\sigma$-coordinate: $\sigma = p/p_{\rm s}$,
where $p_{\rm s}$ is the surface pressure.  This coordinate system is
specifically designed to `follow the terrain' at the bottom.  CAM is
formulated in a more general, {\it hybrid} terrain-following vertical
coordinate, $\eta$: this coordinate system allows the upper part of
the model atmosphere to be represented by $p$-coordinates and the
lower part of the model atmosphere by $\sigma$-coordinates.  To ensure
equitable model inter-comparison, we have set $\eta$ so that $\eta =
\sigma$ throughout the vertical domain in CAM.  In the
$\sigma$-coordinate the vorticity-divergence form of the primitive
equations read:
\begin{subequations}\label{PE CAM}
 \begin{eqnarray}
   \lefteqn{ \frac{\partial{\zeta}}{\partial t}\ \ =\ \bfk\! \cdot\! 
     \bfnabla\! \times\! {\bf n_{\rm{\sigma}}} + \cal{D}_{\zeta}} \\
   \lefteqn{ \frac{\partial{\delta}}{\partial t}\ \ =\ 
     \bfnabla\! \cdot\! {\bf n_{\rm{\sigma}}} - \nabla^2\big(E + \Phi \big) 
     + \cal{D}_{\delta}}\\
   \lefteqn{ \frac{\partial \Phi}{\partial \sigma}\ =\ -\frac{{\cal
         R}T}{p}\frac{\partial p}{\partial \sigma}}\\
   \lefteqn{\frac{\partial p_{\rm s}}{\partial t}\, = \int^{0}_{1}\bfnabla\! 
     \cdot\!\big(p_{\rm s}{\bf v}\big)\, \d\sigma'}\\
   \lefteqn{ \frac{\partial T}{\partial t}\  =\ -\bfv \cdot \bfnabla T -
     \dot{\sigma} \frac{\del T }{ \del \sigma} -\frac{\kappa T \omega}{p} + 
     \frac{\dot{q}}{c_p} + \cal{D}_{T}\, ,} 
\end{eqnarray}
\end{subequations} 
where
\[ {\bf n_{\rm{\sigma}}}\ =\ - \,\big( \zeta + f \big)\,\bfk \times
\bfv - \dot{\sigma}\frac{\partial \bfv}{\partial \sigma} -
\frac{{\cal R}T}{p} \bfnabla p
\]
with $\dot{\sigma}\,\equiv\,\D\sigma / \D t$\, and\, $\D / \D
t\,\equiv\,\del / \del t + \bfv\cdot\bfnabla + \dot{\sigma} \del /
\del \sigma$, the material derivative; ${\cal D}_{\chi}$, where $\chi=
\{\zeta, \delta, T\}$, are given by equation~(\ref{hyper}); and,
$\bfnabla$ here acts on the constant $\sigma$ surfaces.  Note that
these set of equations are slightly different than equations~(\ref{PE
  bob}).

For example, the continuity equation~(\ref{PE CAM}d) comes in the form
of a prognostic equation for surface pressure $p_{\rm s}$ and is
obtained by integrating the continuity equation~(\ref{PE general}c)
from the bottom ($\sigma = 1$) to the top ($\sigma = 0$) surfaces and
using the boundary conditions,
\begin{equation}
\dot{\sigma} = 0 \qquad {\rm{at}} \qquad \sigma = 0, \, 1.
\end{equation}
The pressure vertical velocity, $\omega = \D p / \D t$, is computed
diagnostically from the definition:
\begin{equation}
  \omega\ =\ \sigma\bfv\!\cdot\!\bfnabla p_{\rm s} - 
  \int^{\sigma}_{0}\bfnabla\!\cdot\!\big( p_{\rm s}{\bf v}\big)\, 
  \d\sigma'\, .
\end{equation}
Note, IGCM and CAM employ a vertical discretization scheme described
by \citet{SimmonsBurridge81}.  This vertical finite difference scheme
explicitly conserves mass, total energy and angular
momentum.{\footnote{However, the issue of `hard-wiring' in select
    conservation laws in a numerical scheme is a matter of current
    debate.  For example, an adequately resolved calculation arguably
    does not require a scheme that explicitly enforces global energy
    conservation, which can sometimes lead to unphysical stabilization
    and erroneous results.}

\subsubsection{Finite Volume Core}\label{FV core}

The MITgcm core is widely employed by the Earth's atmospheric and
oceanic communities.  It is highly configurable and is also used in
modeling flows on and in Solar System planets
\citep[e.g.,][]{Kaspi09}.  MITgcm supports both the traditional
hydrostatic and non-hydrostatic formulation of the primitive
equations.  The model, in its traditional formulation, has been
applied in extrasolar planet circulation studies
\citep[e.g.,][]{Showman09,Lewis10}.  In this work, the traditional
hydrostatic formulation is used to ensure equatable comparison.

The primitive equations~(\ref{PE general}) in equally-spaced
$p$-coordinate, in spherical geometry, are solved on a staggered
Arakawa C-grid \citep{Arakawa77} with a second-order finite volume
spatial discretization method \citep[e.g.,][]{Durran99} in the LL grid
and an enstrophy-conserving\footnote{Enstrophy is
  $\frac{1}{2}\zeta^2$.  It is conserved in 2D Euler equation---a 2D,
  barotropic form of equations~(\ref{PE general}) with rigid lid and
  bottom, in the inviscid limit.} scheme \citep{Sadourny75} on the CS
grid---the default configurations of the MITgcm core for the two
grids, respectively.\footnote{We have found embedded in the code two
  additional schemes, which are not described in the official
  documentation, for solving the momentum equation on the CS grid.
  These schemes are not invoked in the default setting, and results
  from detailed tests are presented in a follow-up paper (Polichtchouk
  and Cho, in prep.)} On the LL grid, the grid size approaches zero
near the poles; and, to maintain numerical stability given by the
Courant-Friedrichs-Lewy (CFL) condition \citep[e.g.,][and references
  therein]{Durran99}, an infinitesimal timestep size is required.  To
avoid taking very small timesteps, a fast Fourier transform (FFT)
filter, which smooths out the physically-insignificant grid-scale
waves in the zonal direction, is applied polewards of $45\degr$ at
each timestep.  The problem of grid singularity at the poles in the LL
grid is overcome by the CS grid, which has nearly uniform
grid-spacing.  This grid allows longer timesteps to be taken at a
comparable resolution and eliminates the need for a zonal filter.
However, the CS grid introduces eight special `corner points' (four in
each hemisphere), which lead to an intrinsic wavenumber-4 error in
both the northern and southern hemispheres (see Section~\ref{tc1}).

For the timestepping, a third-order Adams-Bashforth scheme
\citep[e.g.,][]{Durran99} is used.\footnote{Strictly speaking, this
  scheme is not `default' in MITgcm.  However, we have also tested the
  second-order Adams-Bashforth scheme, which is the default, and
  verified that there is no noticeable difference in the results
  between the two schemes.}  The third-order Adams-Bashforth scheme is
more stable, compared with its second-order counterpart, and does not
require a stabilizing parameter to damp the computational mode.  This
scheme is superior to the second-order leapfrog scheme used in the
pseudospectral cores, especially when the second-order scheme is used in
conjunction with the Robert-Asselin filter---as is often the case.

MITgcm supports several dissipation schemes, including harmonic and
biharmonic dissipations, as well as the Shapiro filter
\citep{Shapiro70}.  Because ordinary harmonic dissipation is easy to
implement in both pseudospectral and finite volume cores, Laplacian
dissipation (corresponding to $\mathfrak{p}\! =\! 1$ in
equation~(\ref{hyper})), is applied to the thermodynamic and momentum
equations to control the grid-scale noise in the baroclinic wave test
case.  This approach is similar to the one in \citet{Polvani04} and
ensures that all models solve the same equations, modulo the vertical
coordinate.

Previous extrasolar planet studies with MITgcm have implemented the
Shapiro filter (in CS grid) to control the grid scale oscillations
\citep[e.g.,][]{Showman09,Lewis10}. Since the purpose of the diabatic
test case is to facilitate clear interpretation of outputs from
current hot extrasolar planet studies, we apply the Shapiro filter in
the third test case.\footnote{Note, we have also performed the third
  test case with Laplacian dissipation, as well as with the full range
  of Shapiro filters, for completeness (see Section~\ref{tc3}).}

The Shapiro filter is applied to prognostic variable $\chi$, where
$\chi= \{{\bf v}, \theta\}$, in the zonal and meridional directions.
The discrete form of the Shapiro filter in MITgcm is:
\begin{equation}\label{shap C}
  \widetilde{\chi}_{i,j}\ =\ \left[\,1-\frac{\Delta t}{\tau_{\rm shap}}
    \left\{\frac{1}{8}\Big(\mathsf{F}_{\lambda} + 
      \mathsf{F}_{\phi}\Big)\right\}^{\mathfrak{n}}\,\right]\ \chi_{i,j}\, .\
\end{equation}
Here, $\chi_{i,j}$ is an arbitrary variable at the longitude and
latitude grid points $i$ and $j$, respectively, and is denoted with an
overtilde (i.e., $\widetilde{\chi}_{i,j}$) when smoothed;
$\mathsf{F}_{\lambda}(\cdot)$ and $\mathsf{F}_{\phi}(\cdot)$ are
dimensionless operators operating on $\chi_{i,j}$ such that
\[\mathsf{F}_{\lambda}\big(\chi_{i,j}\big)\ =\ \chi_{i+1,j}-2\chi_{i,j} + 
\chi_{i-1,j}\, ,\]
\[\mathsf{F}_{\phi}\big(\chi_{i,j}\big)\ =\ \chi_{i,j+1}-2\chi_{i,j} + 
\chi_{i,j-1}\, .\
\]
The integer $\mathfrak{n}$ (different from $n$, the total wavenumber)
is the power of the Shapiro filter; $\lambda$ is the longitude;
$\Delta t$ is the timestep size; and, $\tau_{\rm shap}$ is a parameter
which defines the strength of the filter, given $\Delta t$.  In Earth
circulation studies, low power (\,i.e., $\mathfrak{n} = \{2, 4,
6\}$\,) Shapiro filters are generally avoided and are replaced either
by highly scale-selective FFT filters or by less dissipative,
$\mathfrak{n} = \{8, 16\}$, Shapiro filters.  Higher power filters are
chosen to avoid over-dissipating the mid-latitude and tropical waves
\citep[e.g.,][]{Lauritzen11}.  However, we have found that the strong
forcing used in hot extrasolar planet studies generally necessitates
the use of a more dissipative ($\mathfrak{n} \le 6$) filter, for
  the model in its default configuration.  For example, in the
diabatic forcing test case, simulations with MITgcm core at C16
resolution in the CS grid crash for $\mathfrak{n} \geq 8$ Shapiro
filters for all values of $\tau_{\rm shap} \geq
\Delta t$, with $\Delta t$ comparable to those used in pseudospectral
core simulations at similar resolution.\footnote{At this resolution,
  timestep size of typically 5 times smaller than that used in
  pseudospectral cores is required to prevent blow-up in the MITgcm
  core in the default configuration.}

The form of the primitive equations~(\ref{PE general}) solved by
MITgcm in the $p$-coordinate is as follows:
\begin{subequations}\label{PE MITgcm}
 \begin{eqnarray}
   \lefteqn{ \frac{\del\bfv}{\del t}\ \ =\ \
     -\big(\bfv\!\cdot\!\bfnabla\big)\,\bfv - 
     \omega \frac{\del \bfv}{\del p} \ - 
     \bfnabla \Phi - f \bfk \times \bfv + \cal{D}_{{\bf v}}}\\
   \lefteqn{ \frac{\del{\Phi}}{\del p}\ = \ \ -\alpha}\\
   \lefteqn{\bfnabla\! \cdot\! \bfv\ =\ \ -\frac{\del \omega}{\del p}
   }\\
   \lefteqn{\, \frac{\del\theta}{\del t}\ \ =\   
     -\bfv\!\cdot\!\bfnabla\theta - 
     \omega\!\frac{\del \theta}{\del p} + 
     \frac{\theta}{c_p T}\,\dot{q}_{\rm net}+\cal{D}_{\theta}\, , }
 \end{eqnarray}
\end{subequations}
where $\cal{D}_{{\bf v}}$ and $\cal{D}_{\theta}$ represent diffusion.
As discussed above, the diffusion is in the form of
Laplacian dissipation (i.e., $\nu \nabla^2 \chi$,
where $\chi=\{u,v,\theta\}$ and $\nu$ is the constant dissipation
coefficient) in the baroclinic wave test case, while it is in the form
of a ($\mathfrak{n} = 2$) Shapiro filter in the diabatic test case.
Note that, in the latter test case, $\mathfrak{n} = 2$ gives the best
performance in terms of angular momentum conservation in the default
CS grid setting.  Note also that, when solving the primitive equations
in CS grid, a vector-invariant form of equation~(\ref{PE MITgcm}a)
must be used to avoid explicit representation of geometry-dependent
metric terms.  As in the pseudospectral model cores, the equation set
is closed by the ideal gas law and the following boundary conditions:
\begin{equation}\label{BC MIT}
  \omega = 0 \qquad {\rm{at}} \qquad p = 0, \, p_{\rm r}.
\end{equation}
Thus, the upper and lower boundaries act as a solid boundary.

\subsection{Test Cases}\label{tests}

The dynamical cores described in Section~\ref{cores} are subjected to
three tests, which increase in physical complexity.  The test cases
are as follows:
\begin{description}

\item {\it 1})\ \underline{Steady-State} --- assesses the ability of
  the core to maintain a steady-state.  A steady-state is often
  observed in hot extrasolar planet simulations in some parameter
  regimes.  The state in this case is a `neutrally-stable'\footnote{In
    the sense that the jet is stable only in the absence of a
    perturbation -- destabilizing quickly otherwise.},
  zonally-symmetric jet, which is nonlinearly balanced with the
  background temperature.  This is an exact solution to the
  steady-state primitive equations.  In theory, when initialized thus,
  the cores should maintain the state without any change for all
  times, in the absence of external perturbation.  In practice,
  gravity waves and model truncation errors degrade the steady-state
  solution over time \citep{Polichtchouk12}.  A noticeable deviation
  from the initial condition implies the presence of numerical and/or
  programming errors.\\*[-.1cm]

\item {\it 2})\ \underline{Baroclinic Wave} --- assesses the ability
  of the core to faithfully capture the nonlinear evolution of a
  well-studied, three-dimensional flow structure
  \citep[e.g.,][]{Simmons79,Thorncroft93,Polvani04,Jablonowski06,
    Polichtchouk12}.  In contrast to the steady-state case, a small
  perturbation is introduced to the neutrally-stable jet to trigger a
  baroclinic instability, and subsequent evolution over a finite
  duration (20~planetary rotations) is followed.  The magnitude (but
  not the sense and precise location) of the jet is typical of that
  observed in hot extrasolar planet simulations.  The primary aim of
  this test is to clearly expose phase and amplitude errors, which can
  often be obscured by complicated flow evolutions that cannot be
  readily compared with analytic solutions.  Note that, for this
  setup, analytic solutions do not exist.\\*[-.1cm]

\item {\it 3})\ \underline{Diabatic Forcing} --- assesses the
  performance of the core in a setup similar to that used in many hot
  extrasolar giant planet studies in the past
  \cite[e.g.,][]{Showman09,Rauscher10,Thrastarson10}.  In the setup,
  the effect of zonally asymmetric heating from the host star is
  idealized as a simple `Newtonian relaxation' to a prescribed
  temperature distribution in equation~(\ref{PE general}b).  Subject
  to this applied diabatic forcing, the atmosphere is `spun-up' from
  an initial state of rest.  The `strength' of the forcing is
  controlled by the specified day-night temperature gradient and
  characteristic relaxation time which varies in height.  The purpose
  of the test is to elucidate large-scale flow and temperature
  distributions observed in current simulations of tidally
  synchronized extrasolar planets.  In general, the established flow
  and temperature distributions can be variable, depending on the
  forcing and dissipation parameters used
  \citep[e.g.,][]{Cho08a,Thrastarson11}.\\*[-.3cm]
\end{description}

Before presenting the test case results, a brief discussion concerning
convergence is in order.  Throughout the paper we extensively use the
word, `convergence', but take particular care to distinguish three
different types of convergence: {\it numerical}, {\it visual}, and
{\it qualitative}.  In our heuristic definition, numerical convergence
is achieved when a model output is the same up to a specified decimal
precision (e.g., two places), at least at two different spatial
resolutions.  This is the most stringent criterion for convergence and
not easily achieved if the model parameters (e.g., dissipation
coefficient) are different between two resolutions, even for the same
dynamical core.  Visual convergence is less stringent than numerical
convergence and is achieved when plots of the model solutions at two
or more resolutions are nearly visually indistinguishable.
Qualitative convergence is the least stringent definition and is
achieved when the model results at two or more resolutions are similar
in a qualitative sense.  Solutions which differ in phase and amplitude
at a given time, are qualitatively converged if they behave similarly
over a finite time window.  Qualitative convergence can be achieved
within a single core (e.g., at different resolutions) and across
different cores (despite different model parameters).

\section{Results}\label{results}

\subsection{Test Case 1 (TC1): \,Steady-State}\label{tc1}

\subsubsection{TC1 Setup}\label{tc1 setup}

In this test case, a nonlinearly balanced, midlatitude eastward jet is
specified as the initial condition.  The jet is a neutrally-stable
solution to equations~(\ref{PE general}), and is unstable in the
presence of a perturbation.  The setup is identical to the midlatitude
jet setup in \citet{Polichtchouk12}.  We review the setup in
$p$-coordinate only here.  The equivalent setup in $\sigma$-coordinate
can be obtained by using the relation, $p = \sigma p_s$.  The initial
zonal flow $u_0$ is as follows:
\begin{equation}\label{jet_profile}
  u_0(\phi,p)\ =\
  \begin{cases}
    U\,\sin^3\!\big[\pi(\sin^2\phi)\big]\,F(z^*)\, , &
    \hspace*{.3cm} \phi \ge 0\\
    0\hspace*{3.05cm}\, , & \hspace*{.3cm} \phi < 0\, .
  \end{cases}
\end{equation} 
Here, $z^* = -H \log (p/p_{\rm r})$ and
\begin{equation}\label{F_jet}
  F(z^*)\ =\ \frac{1}{2}\left[1 - \tanh^3\left(\frac{z^* - z_0}{\Delta
        z_0}\right)\right]\sin\left(\frac{\pi z^*}{z_{1}}\right)\, .
\end{equation}
The values of the parameters are: $U = 500$~m~s$^{-1}$, $z_{0} =
1823$~km, $z_{1} = 2486$~km, $\Delta z_{0} = 414$~km, $H = 580$~km,
and $p_{\rm r} = 10^5$~Pa ($=10^3~\text{hPa}\approx 1$~bar).

The basic state temperature, $T_0 = T_0(\phi,p)$, is obtained by
combining meridional momentum and hydrostatic balance equations---the
meridional component of equation~(\ref{PE general}a) and
equation~(\ref{PE general}c), respectively:
\begin{equation}\label{gradient wind}
  \frac{\del T_0}{\del \phi}\ =\ -\frac{H}{\cal R}\Big(R_p f + 2u_0 \tan
  \phi\Big)\frac{\del u_0\ }{\del z^*}\, .
\end{equation}
Integrating equation~(\ref{gradient wind}) numerically results in a
temperature distribution that is in gradient-wind balance with the
specified jet (equation~(\ref{jet_profile})).  Here, we have used a
reference temperature of 1500~K as the constant of integration.  The
value is consistent with initial conditions and results of many
extrasolar planet GCM calculations.  In Fig.~\ref{fig1}, the
meridional cross-section of the zonally-symmetric basic state flow
$u_0$ and potential temperature $\theta_0$ (left), as well as the
longitude-latitude map of the relative vorticity $\zeta_0$ at 975~hPa
level (right), are shown.

\begin{figure*}
\begin{center}
  $\begin{array}{cc}
    \hspace*{-0.5cm}
    \includegraphics[height=5cm,width=6.67cm]{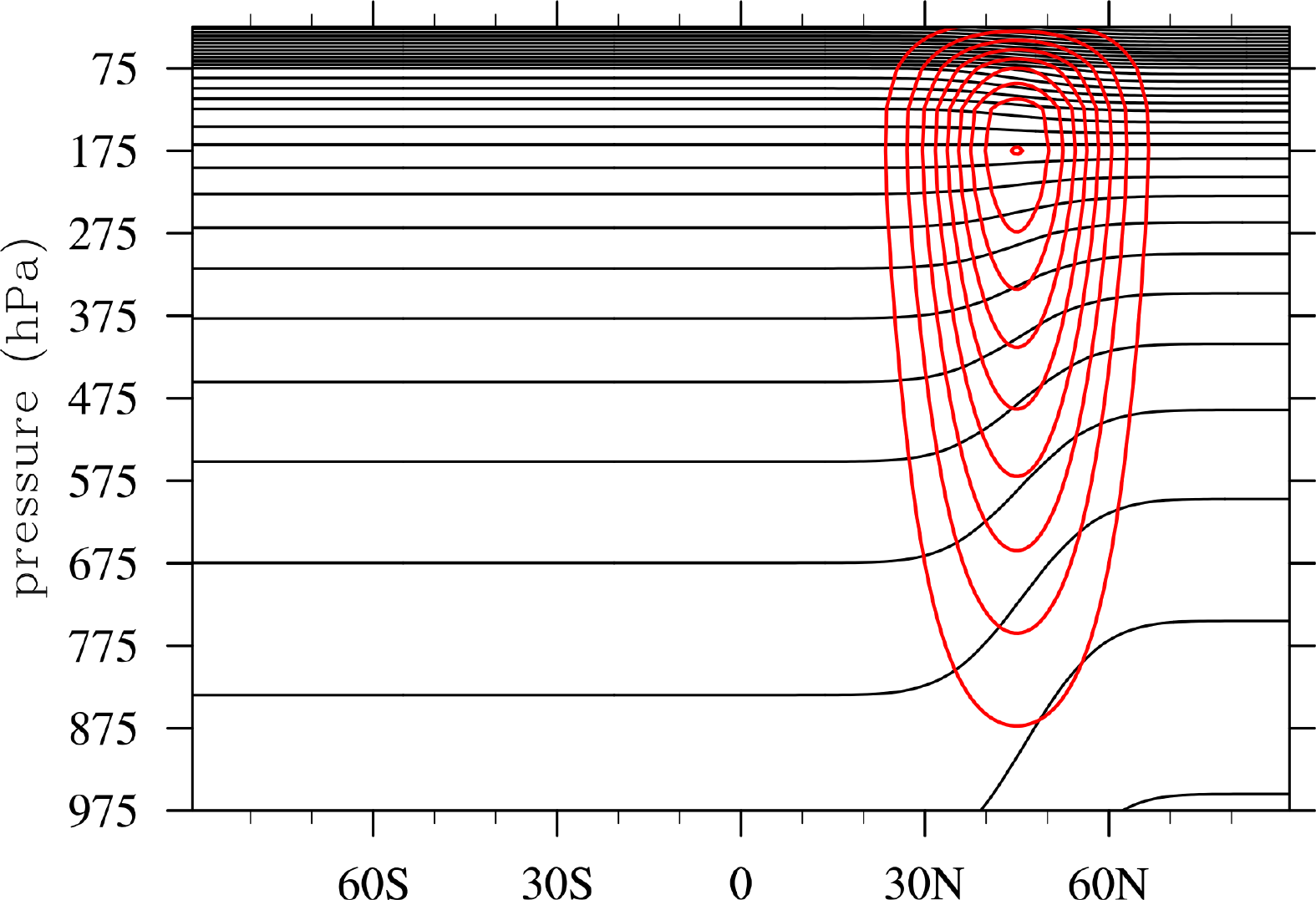}
    \hspace*{0.5cm}
    \includegraphics[height=5.25cm]{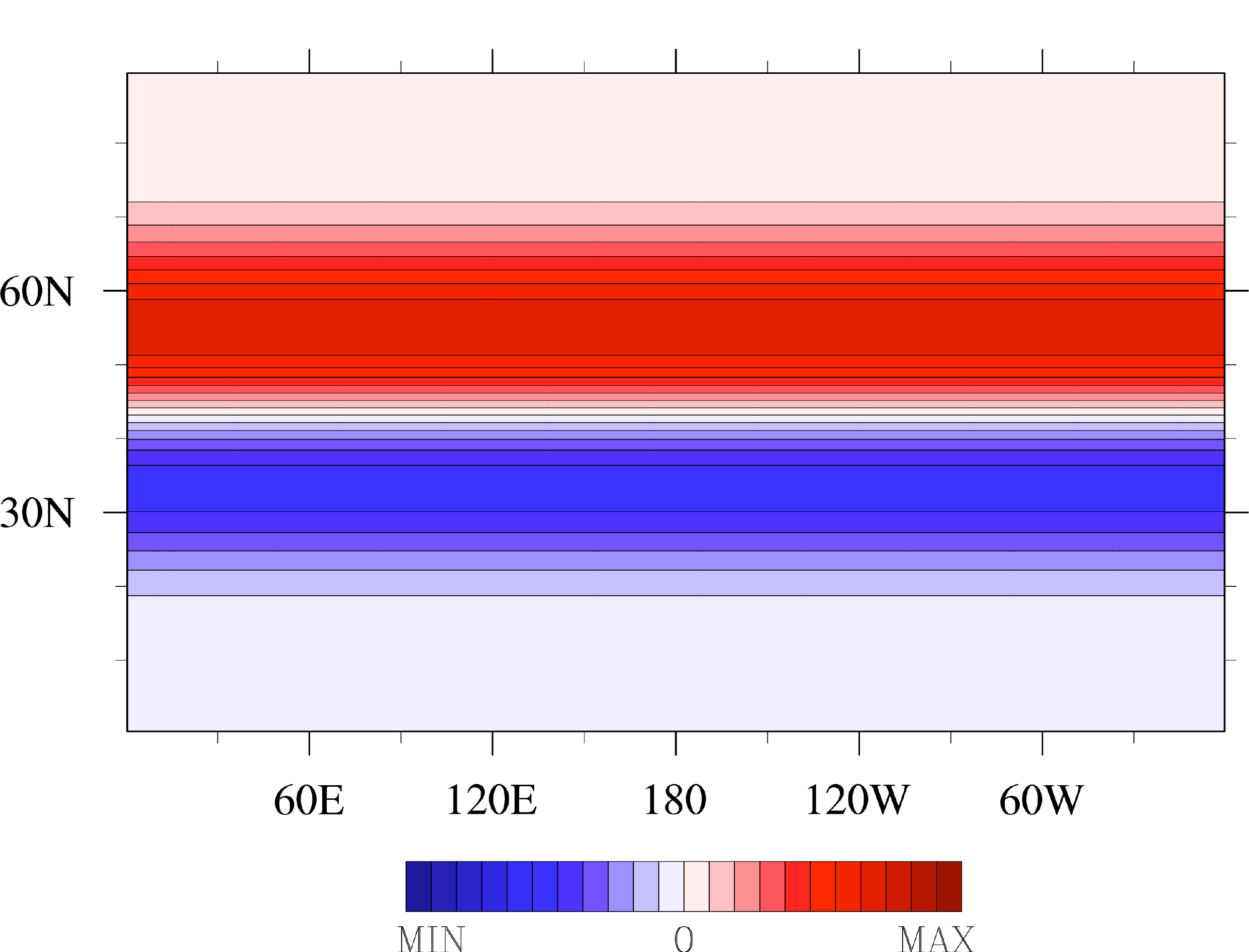}
  \end{array}$
\end{center}
  \caption{\ Left: The basic state zonal wind $u_0$ [m~s$^{-1}$] (red)
    and potential temperature $\theta_0$ [K] (black), as a function of
    latitude and pressure for test case~1, the steady-state test.  The
    contour interval for zonal wind is from 50~m~s$^{-1}$ to
    500~m~s$^{-1}$, in steps of 50~m~s$^{-1}$.  The contour interval
    for potential temperature is from 1400~K to 4400~K, in steps of
    100~K.  The same setup is used as the initial condition in test
    case~2, the baroclinic wave test.\ \ Right:~Basic state relative
    vorticity ($\zeta$) field ~[s$^{-1}$] as a function of longitude
    and latitude in cylindrical-equidistant view, centered on the
    equator, at the 975~hPa ($\approx 975$~mbar) pressure level.  Maximum
    and minimum values are $+5 \times 10^{-7}$~s$^{-1}$ and $-5 \times
    10^{-7}$~s$^{-1}$, respectively, with contour interval of $5
    \times 10^{-8}$~s$^{-1}$. }
  \vspace*{.3cm}
  \label{fig1}
\end{figure*}

Note that, in the above setup, the jet strength is considerably weaker
than in the analogous setup of \citet{Polichtchouk12}.  In this work,
weaker jet amplitude is chosen in order to achieve better numerical
and visual convergence at lower resolutions. Higher resolution is
often required for convergence of high amplitude jets, due to the
stronger ageostrophy associated with high speed jets
\citep[e.g.,][]{Polichtchouk12}.

We emphasize that the initial condition is trivial to set up in all
the models---except for the MITgcm in CS grid configuration.  To
specify the initial zonal wind field in this grid, a MATLAB routine
(supplied with MITgcm) is used to re-grid the wind from the LL grid to
the CS grid.  This re-gridding procedure involves changing the
orientation of the wind velocity vector from the LL grid to the CS
grid by rotating the vector components through grid orientation
angles.  As a result of the procedure, small values of meridional wind
are artificially introduced in the initial wind field; hence, the
initial state becomes slightly less well balanced than that before the
re-gridding.  However, the re-gridding procedure itself does not cause
the destabilization of the jet. We have checked this by re-gridding
the field from the CS grid onto the Gaussian grid: the jet in the
Gaussian grid is stable when the unbalancing effects from the corner
points in the CS grid are removed.

The steady-state case (as well as the other test cases) have been
performed mainly at three different horizontal resolutions. The
vertical domain is resolved by 20 equally spaced $p$ or $\sigma$
levels such that the bottom level midpoint is placed at $p$ = 975~hPa
($\sigma$ = 0.975) and top level midpoint is placed at $p$ = 25~hPa
($\sigma$ = 0.025)\footnote{Note, however, that the bottom interface
  is placed at $p=10^3$~hPa ($\sigma$ = 1) and the top interface at
  $p$ = 0~hPa ($\sigma$ = 0), respectively.}.  The pseudospectral
resolution in the horizontal direction is up to T170 in BOB and up to
T85 in other model cores.  Here, the letter `T' refers to the
triangular truncation and the number refers to the maximum total (as
well as the zonal) wavenumber retained in the spherical harmonic
expansion.  The highest horizontal resolution in MITgcm for the LL
grid is G128 and for the CS grid is C64.\footnote{The C64
      and C16 CS grids have been generated by us with MATLAB routines
      provided by MITgcm support.  However, the default C32 grid (also
      MATLAB generated) comes included with MITgcm---hence, we have
      not generated the C32 grid ourselves.} The `G128' designation refers to
$256\times128$ grid points covering the surface of the sphere.  The
`C64' designation refers to $64\times64$ points covering one of the
six cube faces, for a total of $6\times 64\times 64$ points covering
the entire surface of the sphere.  All the other model specific
parameters needed for reproducing the steady-state test case are
listed in the Appendix, in Tables~\ref{table_a1}--\ref{table_a3}.

The dynamical cores are integrated for 20$\tau$, where $\tau$ is one
planetary rotation (i.e., $2\pi/\Omega$), with no forcing and
dissipation.  Hence, $\bfCalF_{\bf v}\! =\! \bfCalD_{\bf v}\! =\!
{\cal D}_\theta\!  =\!  \dot{q}_{\rm net} = 0$ in equations~(\ref{PE
  general}).  Note that, in the absence of forcing and dissipation,
all dynamical cores should conserve mass, total energy (TE), total
angular momentum (AM) and potential temperature {\it exactly}.  The TE
and AM are defined as:
\begin{eqnarray}
  \qquad {\rm TE} & = & \int_{V}\,\left(\frac{u^2+v^2}{2} + 
    c_{p}T+\Phi\right)\,\d M  \label{TE} \\
  {\rm AM} & = & \int_{V}\,\Big[\big(\Omega R_{\rm p}\cos\phi + 
  u\big)\, R_{\rm p}\cos\phi\,\Big]\ \d M\, , \label{AM}
\end{eqnarray}
where the integral is taken over the volume $V$ of the atmosphere.
Note also, AM is the total absolute angular momentum.

\subsubsection{TC1 Results}\label{tc1 results}

As discussed earlier, all models are expected to maintain the initial
condition (Fig.~\ref{fig1}) because it is an exact solution to
equations~(\ref{PE general}) in the steady state and there are no
external perturbations.  However, in practice the initial state
degrades over time because balance is never perfectly achieved due to
the slight numerical errors generated in the integration of
equation~(\ref{gradient wind}), as well as in the inherent space and
time discretizations.  We quantify the numerical errors via two
$l_{2}$ error norms: the `symmetry' norm and the `degradation' norm
\citep[see, e.g.,][]{Jablonowski06}.  In TC1, these norms are computed
for the zonal wind field $u$.

The symmetry norm assesses the deviation from zonal symmetry (related
to eddy kinetic energy) at each instant.  It is defined:
\begin{eqnarray}\label{L2 symmetry norm}
  \lefteqn{ \hspace*{-.5cm} l_{2}\big[u(t) - 
    \overline{u}(t)\big] = } \nonumber \\
  \lefteqn{ \hspace*{-.5cm}
    \left\{\frac{1}{4\pi}\!\int^{s_{\sub{T}}}_{s_{\sub{B}}}\!
      \int^{\frac{\pi}{2}}_{-\frac{\pi}{2}}\!\int^{2\pi}_{0}\!\!\Big[u(\lambda,
      \phi,s,t) - \overline{u}(\phi, s,t)\Big]^{2}\!
      \cos\phi\,\d\lambda\,\d\phi\,\d s\right\}^{1/2}} \nonumber \\
  \lefteqn{ \hspace*{-.5cm}  
    \approx\!\left\{\frac{\sum_k\sum_j\sum_i\!\big[u(\lambda_i,
        \phi_j,s_k,t) - \overline{u}(\phi_{j}, s_k,t)\big]^{2} w_j\,\Delta s_k}
      {\sum_k\sum_j\sum_i w_j\,\Delta s_k}\right\}^{1/2}\hspace*{-3mm} 
    .} 
\end{eqnarray}
Here, overbar ${\overline{(\cdot)}}$ denotes the zonal average;
indices $i$, $j$, and $k$ are for longitude, latitude, and height,
respectively; $s$ denotes generalized height (and is either $p$ or
$\sigma$ in all the cores); $w_j$ are the Gaussian weights
\citep[e.g.,][]{Canuto88} for the pseudospectral cores or are defined
as $w_j = |\sin\phi_{j+1/2}-\sin\phi_{j-1/2}|$ for MITgcm in LL grid,
where the `half-indices' denote points half way between two
neighboring grid points; and, $\Delta s_k$ are the layer thicknesses.
The degradation norm, on the other hand, assesses the deviation of
zonal average from the zonally-symmetric {\it initial} flow.  It is
defined:
\begin{eqnarray}
  \lefteqn{ l_{2}\big[\overline{u}(t) - \overline{u}(0)\big] = } 
  \nonumber \\
  \lefteqn{ \hspace*{1mm} \left\{\frac{1}{2}\int^{s_{\sub{T}}}_{s_{\sub{B}}}
      \int^{\frac{\pi}{2}}_{-\frac{\pi}{2}}
      \Big[\overline{u}(\phi,s,t) - \overline{u}(\phi,s,0)\Big]^{2}
      \cos\phi\,\d\phi\,\d s\right\}^{1/2} } \nonumber \\
  \lefteqn{ \hspace*{2mm} \approx\left\{\frac{\sum_{k}\sum_{j}
        \big[\overline{u}(\phi_{j},s_{k},t) - 
        \overline{u}(\phi_j,s_k,0)\big]^2 w_{j}\,\Delta s_k}
      {\sum_k\sum_j w_j\,\Delta s_k}\right\}^{1/2}\hspace*{-2mm} .}
\end{eqnarray}
Note, simulation results are interpolated onto a regular LL grid to
compute both $l_2$ norms for MITgcm in CS grid.

In this work, we have found that all pseudospectral cores and MITgcm
in LL grid maintain zonal symmetry to machine precision, at all
resolutions.  The results are not shown, since they are identical to
Fig.~\ref{fig1}.  However, this is not the case for MITgcm in CS grid:
eight special `corner points' (four in the northern hemisphere and
four in the southern hemisphere), where the cube facets meet,
introduce an artificial wavenumber-4 disturbance, quickly degrading
the zonal and temporal symmetry.  This is shown in Fig.~\ref{fig2}.

\begin{figure*}
\begin{center} 
  $\begin{array}{cc}
\hspace*{-0.5cm}
    \includegraphics[height=5cm,width=6.66cm]{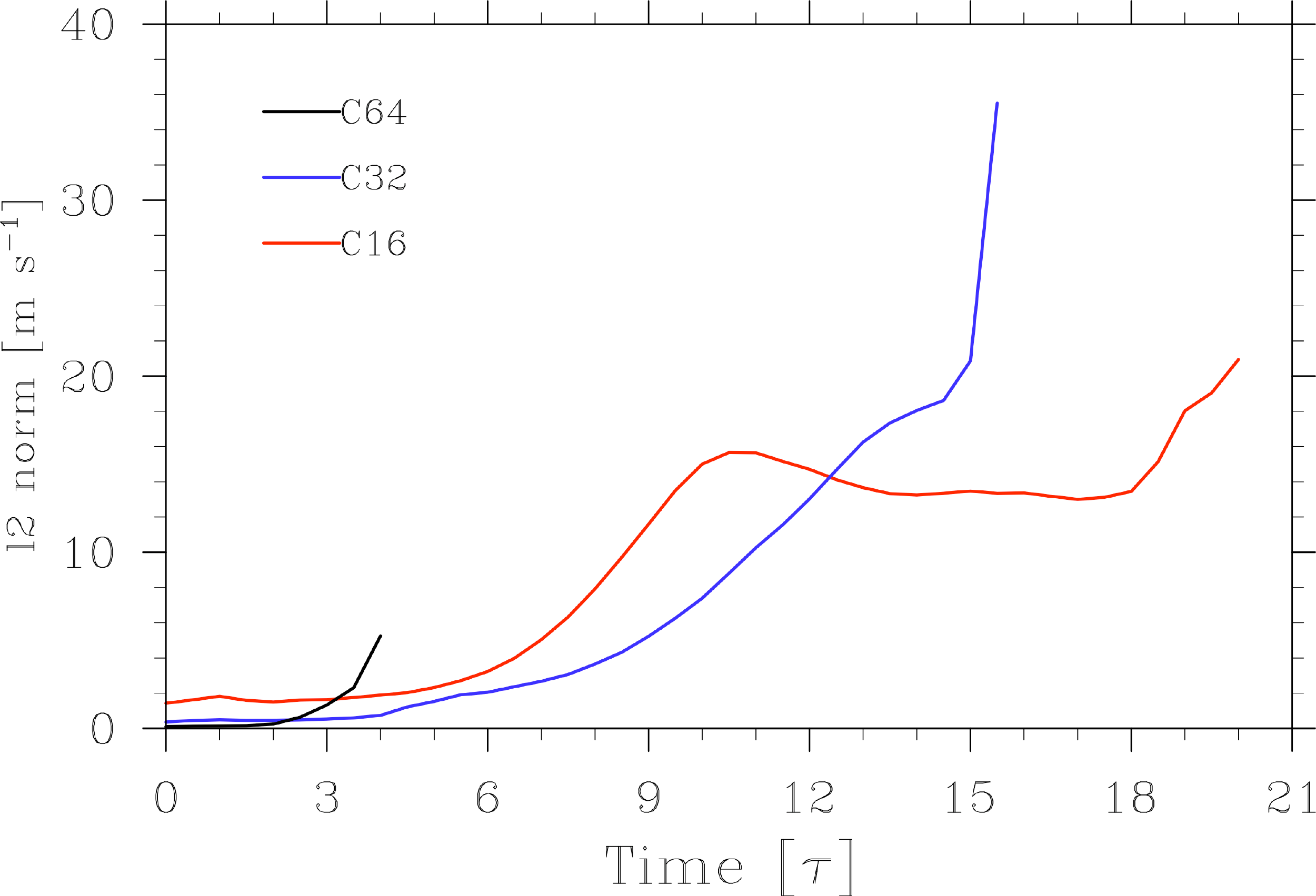}
    \hspace*{0.5cm}
    \includegraphics[height=5.25cm]{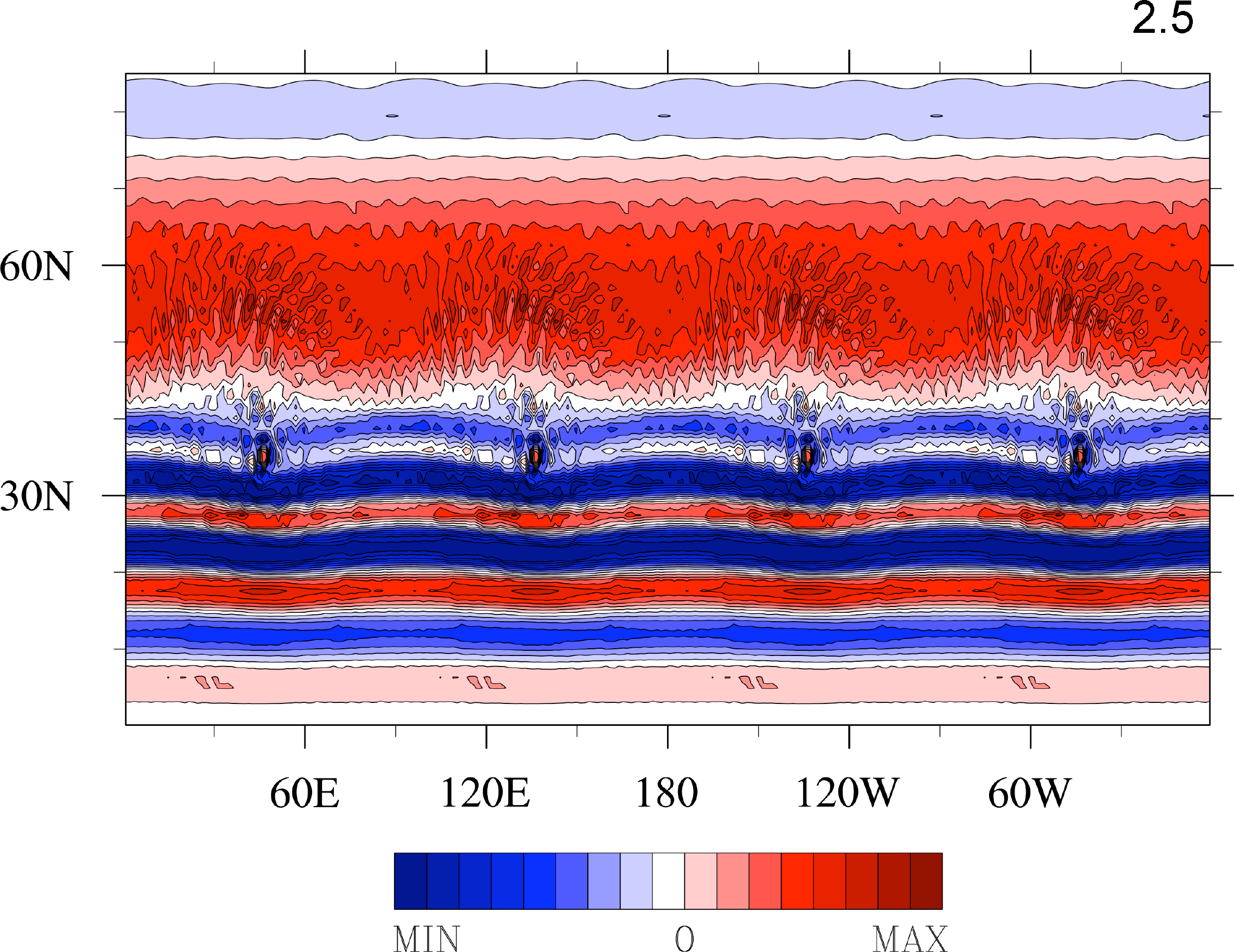}
  \end{array}$
\end{center}
  \caption{\ Left: Symmetry deviation $l_2$-norm of $\big[u(t) -
    \overline{u}(t)\big]$\ \ [m~s$^{-1}$] for MITgcm cubed-sphere (CS)
    steady-state case simulations in the default configuration.
    Three different resolutions (C16, C32, C64) are presented.  Higher
    resolution simulation norms blow-up earlier.\ \ Right: $\zeta$
    field at $t = 2.5\,\tau$, at the 975~hPa level, from the C64
    resolution simulation in the left panel.  Maximum and minimum
    values are $+6 \times 10^{-7}$~s$^{-1}$ and $-6 \times
    10^{-7}$~s$^{-1}$, respectively, with contour interval of $8
    \times 10^{-8}$~s$^{-1}$.  The increase in the $l_2$-norms in the
    left panel are caused by the special corner points, seen in the
    right panel. Note, the norms are not exactly zero initially,
    especially at low resolution. This is due to the errors introduced
    by the re-gridding procedure of $u$ from LL to CS grid.}
  \label{fig2}
  \vspace*{.3cm}
\end{figure*}

In the figure, the left panel shows the $l_2$ symmetry norm.  The
effect of the corners is more pronounced at higher horizontal
resolution, as the grid size near the corners becomes smaller.  The
numerical noise introduced by the corner points causes the higher
resolution simulations to crash earlier---at $t = 15.5 \tau$ and $t =
4.5\, \tau$ at C32 and C64 resolutions, respectively.  Recall that
explicit diffusion is not used in these simulations; but with `enough'
diffusion applied, crashing can be prevented (see, e.g.,
Section~\ref{tc2}).  This is a simple example of when viscosity,
filters, or `fixers' can unintentionally obscure issues in the
numerics and when systematic model intercomparisons can be very
fruitful.  The right panel shows the relative vorticity ($\zeta$)
field from the MITgcm CS simulation at C64 resolution.  The field at
the 975~hPa level is shown in the cylindrical-equidistant projection,
centered on the equator; the time of the simulation is $t = 2.5\,
\tau$.  Only the northern hemisphere is shown.  The numerical noise
from the four special corner points in this hemisphere can clearly be
seen.\footnote{Here, one could argue that this test case (and the next
  one) unfairly favors the Gaussian and LL grids because the jet is
  zonal and passes over the corner points.  However, rotating the grid
  does not fully resolve the adverse effect of the corner points nor
  improves (or reduces the disparity in) the performance of the model
  over a finite duration \citep[see, e.g.,][]{Lauritzen10}.  Note that
  the default setting of the MITgcm in CS grid is the unrotated grid
  configuration.  More importantly, the grid has not been rotated in
  past simulations of extrasolar planets using the MITgcm in CS grid.
  For these reasons, we focus on the unrotated grid configuration in
  this study.}  We have verified that the noise is not due to the
slight imbalance of the flow field introduced by the vector component
rotation, mentioned above (Section~\ref{tc1 setup}).  All other
models, including MITgcm in LL grid, maintain the initial $\zeta$
distribution throughout the duration of simulation ($20\tau$).

\begin{figure*}
  \includegraphics[width=\textwidth]{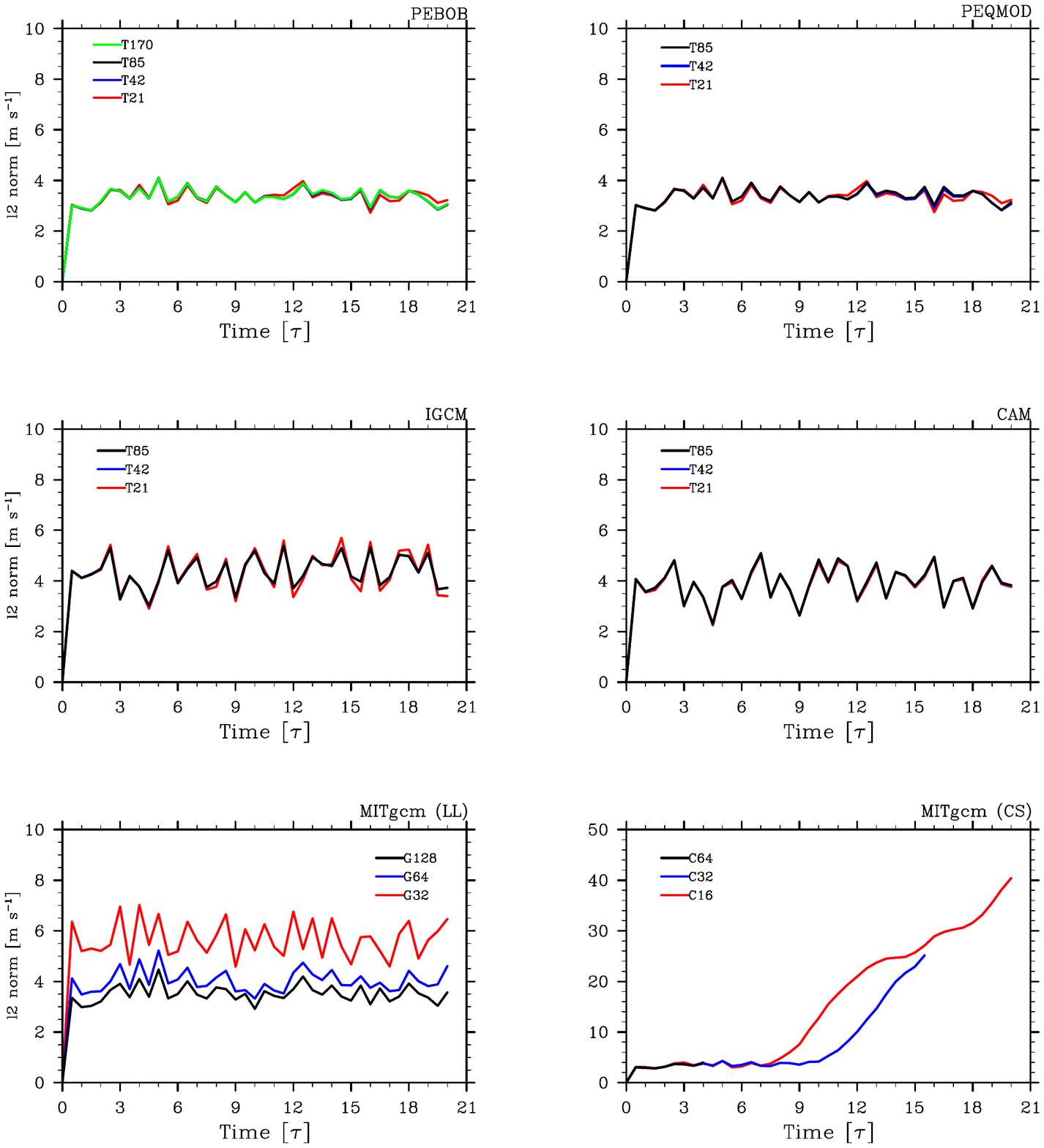}
  \vspace*{-2.5cm}
  \caption{The degradation $l_2$-norm of $\big[\overline{u}(t) -
    \overline{u}(0)\big]$\ \ [m~s$^{-1}$] for all the dynamical cores
    with varying horizontal resolutions.  Note that the vertical scale
    of the plot for the MITgcm in CS grid (bottom right) is five times
    that of the other panels.  In this panel, the three simulations
    are indistinguishable from each other until just before the C64
    simulation crashes, at $t \approx 4.5\, \tau$. }
    \label{fig3}
    \vspace*{.1cm}
\end{figure*}

Fig.~\ref{fig3} shows the $l_2$ degradation norm evolution over the
$20\tau$ duration for all models at varying horizontal resolutions.
The degradation norm presents a more stringent quantification of the
error growth and fluctuation, as well as the intra-core convergence
with resolution, since the deviation is measured against the initial
state.  We first discuss the error growth and fluctuation
characteristics.  This is then followed by a discussion of the
convergence characteristics.

In the figure, the error growths for the pseudospectral cores and the
MITgcm core in LL grid cease, after an initial increase.  The initial
error growth is due to generation of gravity waves, as already
discussed.  The error growth characteristics are identical in BOB and
PEQMOD cores (top row in Fig.~\ref{fig3}), which employ the same
vertical discretization scheme.  The IGCM and CAM cores (middle row in
Fig.~\ref{fig3}) exhibit similar error growth characteristics as BOB
and PEQMOD, but follow much more closely between themselves.  This is
not surprising since IGCM and CAM both use the $\sigma$-coordinate
vertical discretization.\footnote{We remind the reader that the CAM
  core normally uses the more general $\eta$-coordinate \citep[see,
  e.g.,][]{Thrastarson10}, but it has been run in the simpler
  $\sigma$-coordinate to facilitate equatable comparison.}  Note that
both IGCM and CAM show errors saturating at slightly higher levels and
with larger deviations from the saturation level, compared to PEBOB
and PEQMOD.

The MITgcm in the two grids tested, LL and CS grids, show interesting
behavior.  In the LL grid, the core exhibits similar behavior as the
pseudospectral cores---particularly at the higher grid resolutions
(see bottom left panel in Fig.~\ref{fig3}).  At the low grid
resolution (G32), the error saturation level and/or fluctuation
magnitude are larger than in the pseudospectral cores.  However, both
the level and fluctuation magnitude decrease with higher resolution.
In contrast, the core in CS grid exhibits error growth behavior that
is completely different than any of the cores tested (see bottom right
panel in Fig.~\ref{fig3}).  This is expected from the result already
presented in Fig.~\ref{fig2}.  The degradation error in the MITgcm in
CS grid continues to grow with time---again, due to the wavenumber-4
noise from the corner points in the CS grid.  This effect is probably
not so important in simulations of hot extrasolar planet atmospheres,
which are strongly forced non-zonally as well as strongly dissipated.
However, it could have a deleterious influence on steady state and
instability calculations, as demonstrated here (and in the next test
case).

Fig.~\ref{fig3} also shows the convergence characteristics of the
cores.  As can be seen, the pseudospectral calculations are all
visually converged.  CAM calculations are particularly well converged:
the norms for three resolutions tested show essentially no discernible
differences.  On the other hand, the calculations with the MITgcm in
LL grid are not visually converged for resolutions lower than G128,
and this is reflected in the figure (bottom left panel).  Therefore,
these particular calculations are not {\it intra}-model converged.

The above behavior is consistent with the theoretical understanding of
pseudospectral and finite difference and volume methods and past
inter-method comparisons \citep[see, e.g.,][and references
  therein]{Durran99,Boyd00}.  The larger saturation and fluctuation of
the lower grid resolution calculations are likely due to the
second-order accurate finite volume method employed.  For a smooth
flow devoid of shocks or fronts, for example, the resolution of a
pseudospectral simulation is equivalent to an order of magnitude
higher horizontal resolution than in a finite volume/difference
simulation with the same number of degrees of freedom\footnote{This
  also means that, in practice, a finite volume/difference grid should
  not be compared with a Gaussian grid of a pseudospectral method {\it
    with the same number of grid points}, as the latter grid is still
  equivalent to effectively at least three times the resolution of the
  former grid.  The latter point is demonstrated in Fig.~\ref{fig3}
  (cf.~T21 and G32 calculations, for example). Note also that when
  shocks/fronts are present, {\it all} methods have difficulty
  representing the flow accurately, unless specialized treatments
  (available in both pseudospectral and finite difference methods) are
  implemented to specifically deal with the sharp flow structures.}
\cite[e.g.,][]{Canuto88,Durran99,Boyd00,Thrastarson11}; this is
because the order of the pseudospectral method approaches infinity
exponentially fast with increasing resolution.  Note that the MITgcm
in CS grid calculations are neither {\it inter}- nor {\it
  intra}-converged, as the C16 and C32 calculations strongly diverge
after $t \approx 7\,\tau$ and the C64 calculation crashes before this
point at $t \approx 4\,\tau$.

Of all the cores tested, BOB and PEQMOD cores maintain the steady
state solution the best: their norms level off with the smallest mean
value as well as with the smallest root mean square fluctuation from
the mean.  This is partly due to the exclusion of the external gravity
wave mode, a constraint imposed by the `zero vertically-integrated
divergence over the atmosphere' algorithm employed by the two cores
(see Section~\ref{PS cores}).

In terms of convergence, pseudospectral cores are all converged at T42
resolution for this test case.  Differences in the norms, as the
horizontal resolution is increased, is hardly noticeable in these
cores: their solutions are visually converged.  As already mentioned,
the CAM solutions show remarkably little difference at different
resolutions: they are numerically converged.  {\it Numerical}
convergence is achieved in BOB at T85 resolution; the norms for T85
and T170 resolutions match exactly up to the second decimal place.
Again, this is expected, given the exponential convergence property of
the pseudospectral method.  In contrast, the MITgcm calculations are not
visually converged in both LL and CS grid configurations.

In summary, apart from the MITgcm in CS grid, the steady-state
condition is well maintained throughout the duration of the
calculations by all of the cores.  Hence, these calculations are
qualitatively inter-model converged for this test case at the
resolutions considered.  As additional measures of convergence, we
have verified that these cores conserve the total initial energy
(equation~(\ref{TE})) and total angular momentum
(equation~(\ref{AM})).  The values of the two quantities are
$2.3\times 10^{28}$~J and $1.8\times 10^{32}$~J~s, respectively, for
this test case.  These values are maintained throughout the
integration to within 0.02 percent (except, of course, for the MITgcm
in CS grid).  At this point, one may be tempted to down-play the
differences between the model cores reported here---particularly in
the pseudospectral cores.  However, we caution that even such small
discrepancies can---and in practice do---lead to non-trivial
differences in the model outputs, if the problem is more complex or
requires high spatio-temporal accuracy (e.g., instability and
transition to turbulence).

\subsection{Test Case 2 (TC2):\ Baroclinic Wave}\label{tc2}

\subsubsection{TC2 Setup}\label{tc2 setup}

In this case, an instability is initiated in the neutrally-stable
state of Section~\ref{tc1} to generate an nonlinearly evolving
baroclinic wave.  The instability is triggered by perturbing the
initial temperature $T_0$ with a heat bump $T'$ at all pressure
levels, where
\begin{equation}\label{Tprime}
  T'(\lambda,\phi)\ =\ \mbox{sech}^2\big(3\,\lambda\big)\, 
  \mbox{sech}^2\left[6\left(\phi-\frac{\pi}{4}\right)\right]\, .
\end{equation}
Once the instability ensues, the flow is allowed to evolve freely
thereafter for 20\,$\tau$.  It is important that exactly this
perturbation is used, when attempting to reproduce the results here.
This is because, while the flow is expected to asymptotically reach
qualitatively the same state, the early-time evolution is different
for a different perturbation.  It is then difficult to delineate the
source of the variations in the subsequent evolution---whether the
variations are due to physically different modes being excited or to
numerical inaccuracies.

The instability leads to a rapid development of sharp fronts in few
planetary rotations (i.e., few $\tau$'s), and we can no longer
integrate the inviscid equations (as in TC1) since there is a rapid
build up of energy in small scales.  This case is arguably more
`realistic' than the steady-state case, in the sense that explicit
viscosity must be used---as in most long-duration simulations
involving complex flows.

To make the comparison easier, we choose to implement in this test
case a Laplacian dissipation operator ($\nabla^2$) in all
the cores, even though more scale-selective, higher order,
hyperdiffusion operators (e.g., $\mathfrak{p}\ge 2$ in
equation~(\ref{hyper})) are almost always used in pseudospectral
calculations.  Although hyperdiffusion operators acting on vorticity
and divergence fields are common in pseudospectral cores, they are
less common in finite volume cores because they are more difficult to
implement in the finite volume discretization scheme.  In the latter
type of cores, alternative strategies are used to effect high
scale-selectivity.  As discussed in Section~\ref{FV core}, in addition
to the harmonic (second-order) and biharmonic (fourth-order)
diffusion, the MITgcm also supports the Shapiro filter.

As in \citet{Polvani04} and \citet{Polichtchouk12}, the same value of
dissipation coefficient ($\nu_2 = 2\times 10^7$~m$^2$~s$^{-1}$) is
used for all the resolutions in TC2.  The usual practice is to
adjust---or tune---the value for each resolution, problem and model
(see, e.g., discussions in \citet{Thrastarson11} and
\citet{Polichtchouk12}).  However, $\nu_2$ is not adjusted in this
case so that each mode, up to the truncation wavenumber, experiences
the same amount of dissipation, regardless of the resolution.  For
example, the dissipation time at the T21 truncation scale for
HD209458b corresponding to the above value of $\nu_2$ is: $\tau_{\rm
  d} = 3.58\,\tau$.  In comparison, current flow modeling studies of
hot giant extrasolar planets employ a much shorter damping time of
$\tau_{\rm d} \sim 0.02\,\tau$ \citep[e.g.,][]{Rauscher10,Heng11};
hence, these simulations are more dissipative than the ones in this
study.  However, the damping time used is still generally shorter than
that used in the previous, similar study by
\citet{Polichtchouk12}---again, to allow a more equatable comparison
between the different cores to be performed.

The highest horizontal resolutions investigated in this test case are
the same as in the steady-state test case (TC1). The resolution
specifications and other parameters needed for reproducing this test
case are listed in the Appendix,
Tables~\ref{table_a4}--\ref{table_a6}.  Note, unlike in TC1, the true
solution to the primitive equations is unknown for this test case.

\subsubsection{TC2 Results}\label{tc2 results}

Fig.~\ref{fig4} shows the evolution of $\zeta$ at the $p = 975$~hPa
surface from a simulation with BOB at T170L20 (i.e., T170 horizontal
resolution with 20 vertical levels) resolution, for $t = 6\,\tau$ to
$t = 18\,\tau$.  The 975~hPa pressure surface is chosen because the
maximum eddy activity of the unstable evolution occurs near the lower
boundary \citep[e.g.,][]{Polichtchouk12}.  In the evolution, the
perturbed jet undergoes a period of linear growth ($t \ll 9\,\tau$),
when the most unstable mode (mode~3--4) emerges.  By $t \approx
14\,\tau$ the evolution is well in its nonlinear stage, characterized
by the exponential growth of eddy kinetic energy and wave breaking.
The $\zeta$ perturbation exhibits a distinct northwest-southeast tilt
on the poleward side of the jet and southwest-northeast tilt on the
equatorward side of the jet.  Near $t = 18\,\tau$, the eddy kinetic
energy reaches the maximum value of the simulation and the barotropic
decay cycle, in which eddy kinetic energy is returned back to the mean
flow, ensues for $t \gtrsim 20\,\tau$.  In this period, the cyclones
(areas of positive $\zeta$ anomalies, shown in red in the figure) that
have emerged from the wave breaking, start interacting and advance
poleward \citep[e.g.,][]{Cho08b,Polichtchouk12}.

\begin{figure*}
\begin{center}
  \includegraphics[width=\textwidth]{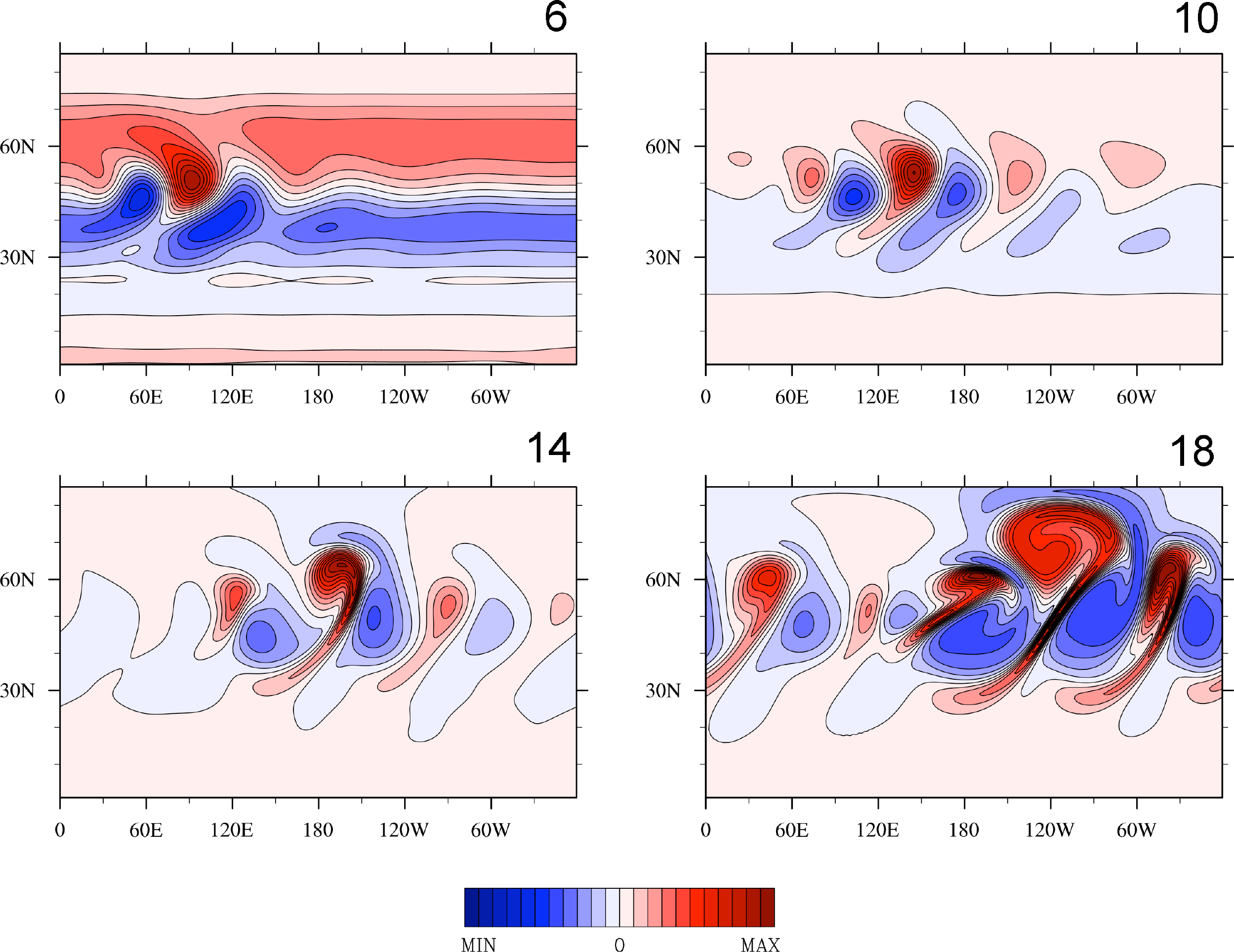}
\end{center}
  \vspace*{.5cm}
  \caption{\ Relative vorticity ($\zeta$) field from T170L20 run with
    BOB in cylindrical-equidistant view, centered on the equator.  The
    fields are shown at the 975~hPa pressure level for $t = 6\,\tau$
    to $t = 18\,\tau$.  Maximum and minimum values are: $\pm 1 \times
    10^{-6}$~s$^{-1}$ ($t = 6\,\tau$); $\pm 5 \times 10^{-6}$~s$^{-1}$
    ($t = 10\,\tau$); and, $\pm 2 \times 10^{-5}$~s$^{-1}$ ($t =
    14\,\tau$ and $t = 18\,\tau$).  The contour intervals are,
    respectively, $1 \times 10^{-7}$~s$^{-1}$, $5 \times
    10^{-7}$~s$^{-1}$ and $2 \times 10^{-6}$~s$^{-1}$.  Note the
    large, an order of magnitude, change in the amplitude of $\zeta$
    during the evolution---as well as the formation of sharp fronts
    and coherent vortices, particularly at $t = 14\,\tau$ and $t =
    18\,\tau$.}
  \label{fig4}
\end{figure*}

The evolution presented in Fig.~\ref{fig4} is the high resolution
`reference solution' for the BOB core.  For the other pseudospectral
cores and the MITgcm core in LL and CS grids, the reference solution
is computed at T85L20, G128L20, and C64L20 resolutions, respectively.
In addition to these solutions, the outputs from the other model cores
may be compared with the T170L20 reference solution obtained with the
BOB core.  In principle, since all cores solve the same equations (and
boundary conditions), the high resolution reference solution computed
with one of the cores should serve as a reference solution for all the
models.  However, in practice, there is a danger in using a single
model core to determine the reference solution in problems involving
unstable states, as noted by \citet{Jablonowski06}.  This is due to
the differences in how various model cores handle geostrophic
adjustment.

Fig.~\ref{fig5} compares the solutions of all the dynamical cores at
the T85L20, G128L20, and C64L20 resolutions at $t = 10\,\tau$
(cf.~upper right frame in Fig.~\ref{fig4}).  As in TC1, the
calculations with IGCM and CAM look nearly identical.  The same is
true for BOB and PEQMOD calculations, although the amplitude of
vorticity anomalies in PEQMOD is somewhat stronger than in BOB.
Comparing with IGCM and CAM, the amplitude in PEQMOD is noticeably
stronger.  In general, we have found that the vorticity anomalies are
stronger in the $p$-coordinate pseudospectral cores than in
$\sigma$-coordinate pseudospectral cores.  In contrast, the phases are
impressively similar among all the pseudospectral cores.

\begin{figure*}
  \includegraphics[width=\textwidth]{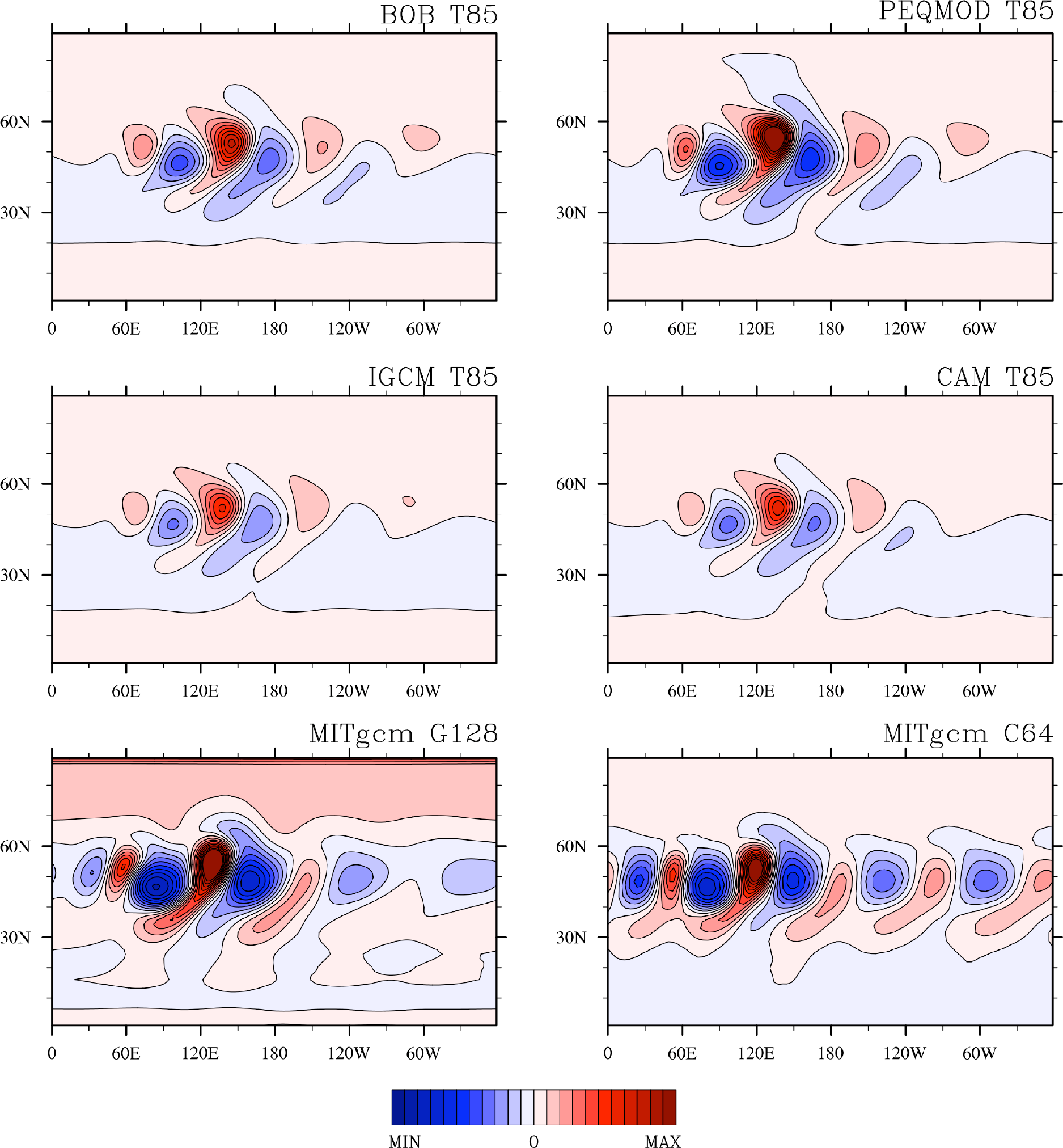}
  \vspace*{.5cm}
  \caption{\ Cylindrical equidistant view, centered on the equator, of
    $\zeta$ field at $t = 10\,\tau$ from different model cores.  The
    resolution is the highest tested in all the cores, except in BOB;
    it is the second highest.  The fields shown are from the bottom
    vertical level ($\sim$ 975~hPa).  Maximum and minimum values for
    all the cores are $\pm 5\times 10^{-6}$~s$^{-1}$, with contour
    interval $5 \times 10^{-7}$~s$^{-1}$. These fields are to be
    compared with each other, as well as with that at $t = 10\,\tau$
    in Fig.~\ref{fig4}.}
  \label{fig5}
\end{figure*}

There are, however, considerable differences in both amplitude and
phase between solutions with MITgcm and pseudospectral cores.  Compare
the overall vorticity fields, and especially the magnitude of the
$\zeta$ anomalies in the fields.  This is partly caused by $\zeta$ not
being a prognostic variable in the MITgcm core: $u$ and $v$ fields are
evolved, and the calculation of $\zeta$ from these fields introduces
some errors.  If the potential temperature ($\theta$) field is
compared instead of the $\zeta$ field, the MITgcm LL grid solution
resembles the corresponding PEQMOD solution more closely (not shown).
However, even using the $\theta$ field, the MITgcm CS grid solution
differs significantly from the pseudospectral core solutions.  This
difference---between the two MITgcm solutions, in LL and CS grids---is
revealing (cf.~two bottom panels in Fig.~\ref{fig5}).  Note that the
only difference between the two calculations is the grid (and the use
of a high-wavenumber zonal filter in the LL grid calculation).

The solution from MITgcm in CS grid clearly exhibits a different
unstable mode structure than that of the LL grid solution at $t =
10\,\tau$ (i.e., $\sim\! 5$ for the CS grid compared with $\sim\! 4$
for the LL grid).  The difference is due to the corners in the CS
grid, which provide an additional source of perturbation. We have
verified this by running a MITgcm in CS grid baroclinic wave
simulation without perturbing the background temperature $T_0$ by $T'$
(see equation~\ref{Tprime}).  In this case a clear mode-4 structure
associated with the corner points (different from those produced in
pseudospectral core calculations shown in Fig.~\ref{fig5}) dominates
the evolution throughout.

To assess the convergence (with resolution) characteristics of a
model, we compute the $l_2$ relative vorticity norm at the lowest
model layer (i.e., the $\sim$975~hPa pressure surface).  The norm is
defined as follows:
\begin{eqnarray}
  \lefteqn{\ \ l_2\big[\zeta(s_{20})\big]\ = } \nonumber\\ 
  \lefteqn{\quad \left\{\frac{1}{4\pi}
      \int^{2\pi}_0\int^{\frac{\pi}{2}}_{-\frac{\pi}{2}} 
      \Big[\zeta(\lambda,\phi,s_{20},t)\Big]^2
      \cos\phi\,\d\phi\,\d\lambda \right\}^{1/2} } \nonumber \\
  \lefteqn{ \qquad \approx\ 
    \left\{\frac{\sum_i\sum_j\big[u(\lambda_i,\phi_j,s_{20},t)\big]^{2}w_j}
      {\sum_i\sum_jw_j}\right\}^{1/2},} 
\end{eqnarray}
where $s_{20}$ is the lowest model layer in $p$- or
$\sigma$-coordinate and the sums are taken over all
$(\lambda_i,\phi_j)$ points on the sphere.  The integration weights
are defined as in equation~(\ref{L2 symmetry norm}).  After
calculating the $l_2[\zeta(s_{20})]$-norm for each resolution,
the highest resolution $l_2$-norm is subtracted from lower resolution
$l_2$-norms within the same model to assess model core convergence.

The differences between $l_2[\zeta(s_{20})]$-norms within the same
core are shown in Fig.~\ref{fig6}.  It is clear that none of the cores
are (numerically or visually) converged at the second lowest
resolution---T42, G64 and C32: the curves all deviate from zero
(cf. in the upper left panel the `T85$-$T170' curve, which shows visual
convergence).  The non-convergence is particularly apparent after the
baroclinic instability enters the fully nonlinear growth stage ($t
\gtrsim 10\,\tau$).  Even at T85 resolution, BOB is only visually (not
numerically) converged.  This can be verified by comparing the top
right panel of Fig.~\ref{fig4} to the top left panel of
Fig.~\ref{fig5}: the plots in the two panels are very close to each
other but not identical.

\begin{figure*}
  \vspace*{-2.25cm}
  \includegraphics[height=22cm,width=18cm]{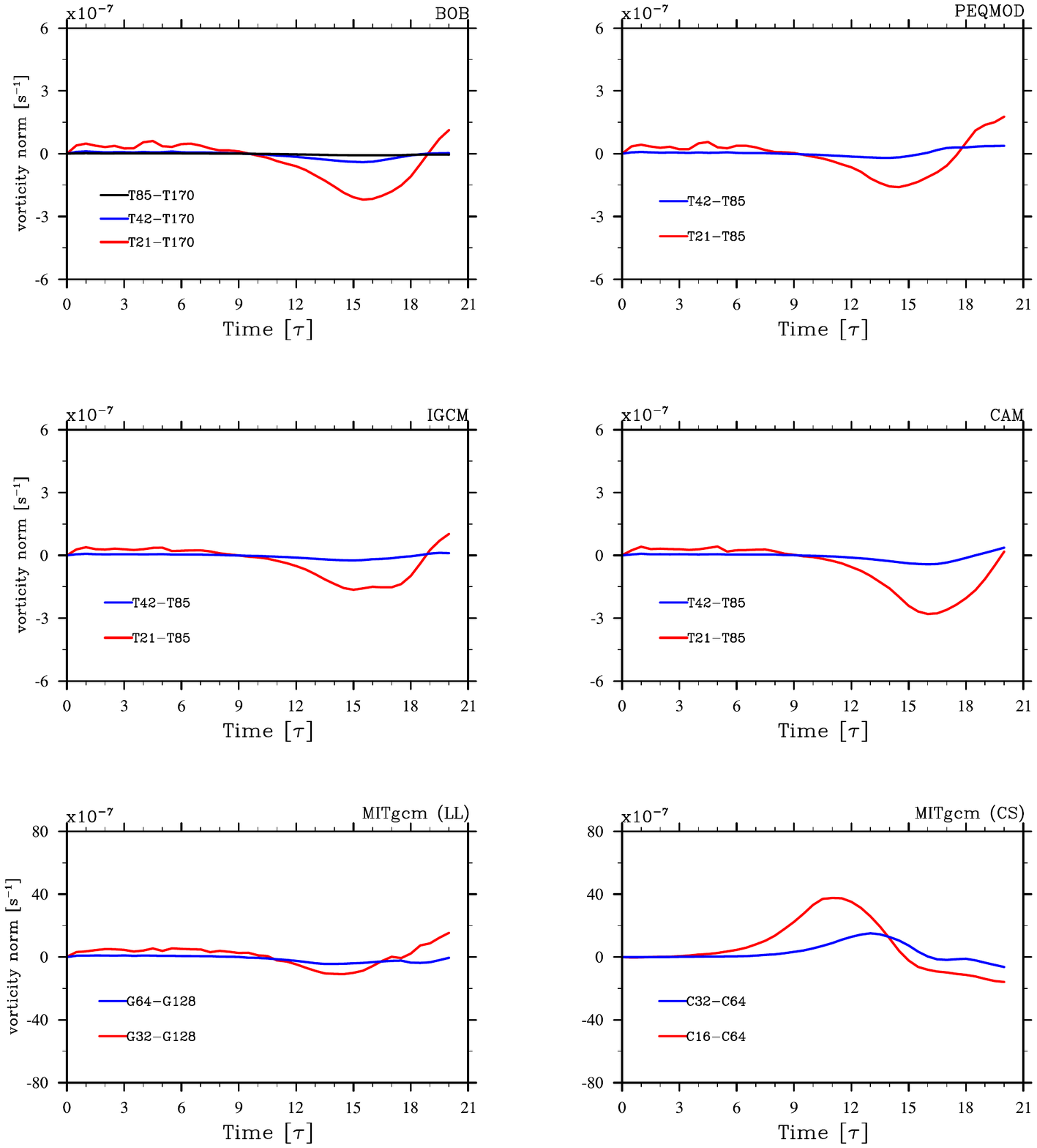}
  \vspace*{-2.75cm}  
  \caption{\ Differences of the root mean square $l_2$ vorticity norm
    [s$^{-1}$] between high resolution reference solution and lower
    resolution solutions within the same model core.  The T85 BOB
    simulation is well converged.  No other cores are `intra-model'
    converged.  Note, the $y$-scale in the two plots of the bottom row
    for the MITgcm core is nearly an order of magnitude larger than in
    the other plots. }
  \label{fig6}
  \vspace*{.2cm}
\end{figure*}

Note that the $y$-scale in the MITgcm plots in Fig.~\ref{fig6} (as
well as in Fig.~\ref{fig7}) is an order of magnitude larger than in
the corresponding pseudospectral core plots.  This suggests that the
apparent qualitative inter-convergence of the MITgcm core in LL grid
at G128L20 resolution (seen in Fig.~\ref{fig5}) is suspect.  We have
verified that large differences in the $l_2$-norm in MITgcm are not
caused by the errors introduced in the calculation of $\zeta$.  For
example, the $l_2$-norm of the surface $u$-field behaves in a similar
way and visual convergence is clearly not achieved in the MITgcm core
at the highest resolution tested in this work.

\begin{figure*}
  \vspace*{-2.25cm}
  \includegraphics[height=22cm,width=18cm]{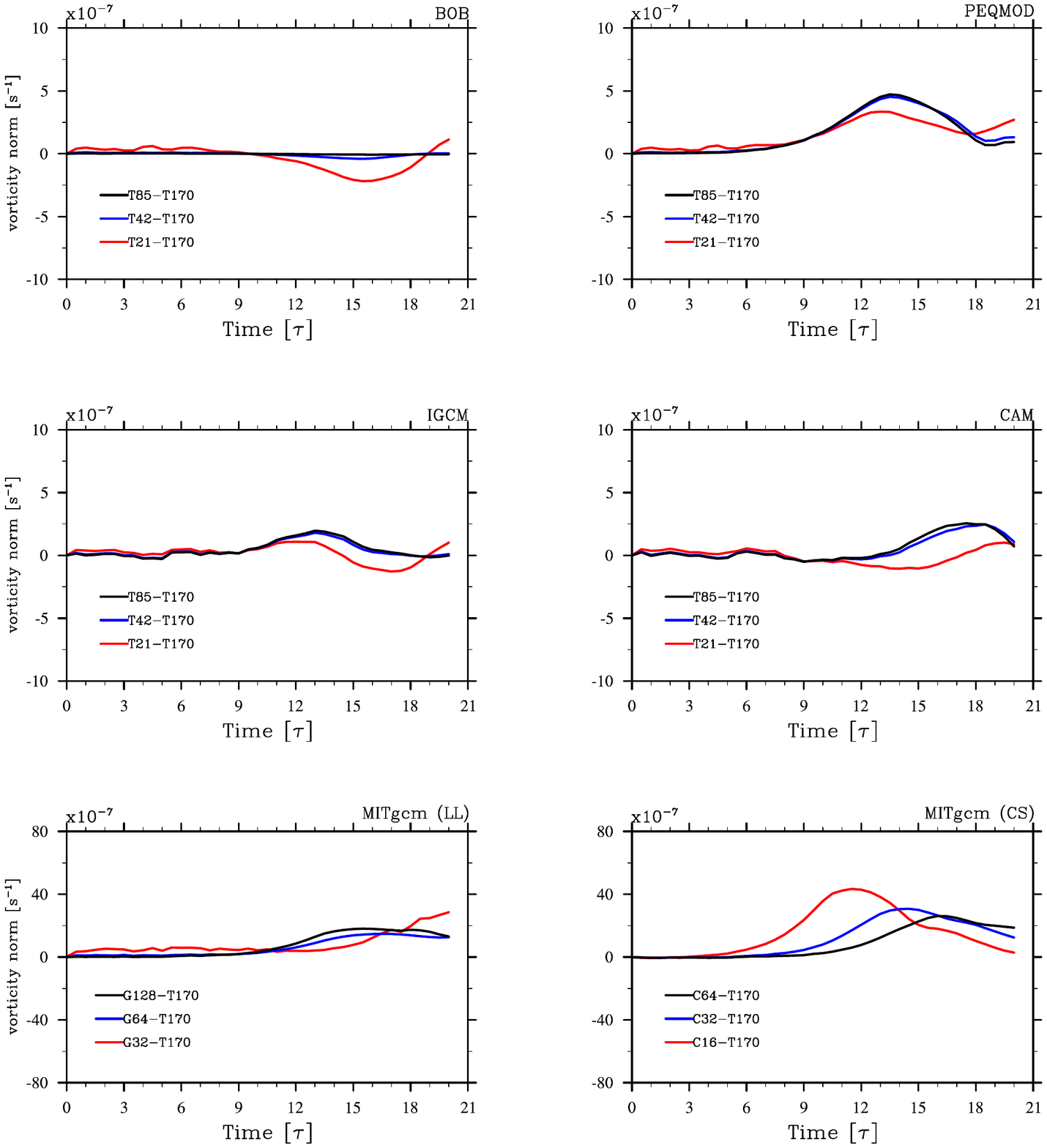}
  \vspace*{-2.5cm}
  \caption{\ Differences of the root mean square $l_2$ vorticity norm
    [s$^{-1}$] between T170L20 reference solution for BOB and
    solutions with other dynamical cores at various resolutions.  The
    scales are same as in Fig.~\ref{fig6}.  The T42 simulations in all
    the pseudospectral cores appears to be marginally
    `intra-converged'.  MITgcm core in both LL and CS grids are not
    converged, particularly in the latter grid.}
  \label{fig7}
  \vspace*{.4cm}
\end{figure*}

Fig.~\ref{fig7} shows the differences between
$l_2[\zeta(s_{20})]$-norms of a given core and the T170
$l_2[\zeta(s_{20})]$-norm of BOB.  Hence, here we are treating the
T170 calculation as a `reference solution'---as if it were the
`correct solution'.  Recall that ideally calculations by each core
should be compared with T170 (or higher) resolution calculation of the
same core.  For technical reasons, this is not feasible in all the
cores.  However, this comparison is still useful and provides some
insights.  The figure clearly demonstrates that solutions of other
model cores are not visually converged to BOB's high-resolution
solution.  In all model solutions, the norms are small until
$t\!\sim\! 9\,\tau$, but then increase markedly once the evolution
enters the fully nonlinear stage, when the wave begins to develop
sharp fronts.  This is expected, given the behavior of baroclinically
unstable evolution reported in \citet{Polichtchouk12}.

Note that the evolution of the baroclinic wave growth in this set of
simulations is considerably retarded by the stronger explicit
dissipation, compared to that in \citet{Polichtchouk12} (recall that
Laplacian dissipation, rather than superdissipation,
is employed here).  Moreover, the initial jet's maximum speed is
weaker here than that in \citet{Polichtchouk12}, leading to a smaller
growth rate for the baroclinic instability.  \citet{Polichtchouk12}
have demonstrated that baroclinic instability in a similar situation
is only marginally captured with T85 resolution for a 1000~m~s$^{-1}$
jet.  Consistent with that study, all pseudospectral cores are
visually converged at resolution T85 in this study, as expected for a
500~m~s$^{-1}$ jet.  The more stringent resolution criterion for
convergence in \citet{Polichtchouk12} is due to the stronger
ageostrophy present in a faster jet, as already discussed in
Section~\ref{tc1 setup}.  Similarly, if the initial jet amplitude had
been greater than 1000~m~s$^{-1}$, the resolution at which convergence
would be achieved is expected to be correspondingly
higher.\footnote{Note, speeds greater than 2000~m~s$^{-1}$ are often
  produced in many hot-Jupiter simulations.  Some simulations produce
  speeds which are much greater than this.}

Interestingly, even with the application of Laplacian dissipation,
wchich is a strong dissipation, the global energy is conserved to
within 0.1 percent in all the model cores throughout the integration
($t =20\, \tau$).  Global angular momentum is conserved to within 0.02
percent in all the model cores throughout the integration.  Note that
this is for Laplacian dissipation only, as only this dissipation is
used in TC2.  These conservation properties should be compared with
the corresponding ones in the next test case.

\subsection{Test Case 3 (TC3):\ Diabatic Forcing}\label{tc3}

\subsubsection{TC3 Setup}\label{tc3 setup}

No `physical processes' (e.g., net heating, wave drag, convection)
were specified in the setup of TC1 and TC2, if explicit dissipation is
not considered to be representing `turbulent viscosity' in the latter
test case.  Most of these processes are as yet poorly constrained by
observations or unobtainable from first principles for extrasolar
planets \citep[see, e.g., discussion in][]{Cho08a, Cho08b,Showman10}.
Two such processes are irradiation from the host star and radiative
cooling in the atmosphere of the planet.  These processes are
currently represented essentially in all extrasolar planet atmosphere
simulations in a highly idealized way.  For example, Newtonian
relaxation parameterization to a prescribed `equilibrium temperature'
is often used to crudely represent the combined thermal forcing
\citep[see,
e.g.,][]{Cooper05,Showman08a,Menou09,Rauscher10,Thrastarson10,
  Thrastarson11,Heng11}.  Despite the crudeness, here we also use the
parameterization---given its simplicity and common usage in past
works.  The idea is to be reasonably close to past simulations, while
facilitating reproducibility of the present work and clean comparisons
with future work.

In the Newtonian relaxation parameterization, the source term
($\dot{q}/c_p$) in the thermodynamic equation (equations~(\ref{PE
  bob}e), (\ref{PE CAM}e) or (\ref{PE MITgcm}d)) is specified as
\begin{equation}
  \frac{\dot{q}}{c_p}\ =\ -\frac{1}{\tau_{\rm th}}\big(T - T_e \big)\, .
\end{equation}
Here, in its general form, $T_e = T_e (\lambda,\phi,s,t)$ is the
equilibrium temperature and $\tau_{\rm th} = \tau_{\rm
  th}(\lambda,\phi,s,t)$ is the thermal relaxation time.  Both $T_e$
and $\tau_{\rm th}$ distributions are currently not well known, both
in space and in time.  In many studies, very short relaxation times
($\,\ll\! 1$~hour) and large equilibrium temperature gradients across
the day-night terminator ($\,\approx\!  1000$~K) are specified
\citep[e.g.,][]{Showman08a, Rauscher10,Thrastarson10}.  Such a
condition constitutes an `extreme forcing' on the
dynamics---especially in simulations started from rest state, spun up
to a strongly unbalanced state
\citep{Cho08a,Thrastarson11,Polichtchouk12}: the Solar System planets
are characterized by comparatively much longer $\tau_{\rm th}$ and
much smaller $T_e$ gradient.

In this test case, $T_e$ is height-independent (i.e., $\partial
T_e/\partial s = 0$) and both $T_e$ and $\tau_{\rm th}$ are steady
(i.e., $\partial\,\{T_e,\tau_{\rm th}\}/\partial t = 0$).  In general,
both $T_e$ and $\tau_{\rm th}$ are complicated functions of space and
time \citep{Cho08a,Showman09}.  However, in keeping with the overall
aim of this work, we choose a setup which is at once easily
describable and easily implementable in all models.  Here, we choose
$T_e$ to be as in \citet{Thrastarson11}:
\begin{equation}
  T_e\ =\ T_m + \Delta T_e \cos \phi \cos\lambda\,\ ,
\end{equation} 
where $T_m = (T_D + T_N)/2$ and $\Delta T_e = (T_D - T_N)/2$ with
$T_D= 1900$~K and $T_N = 900$~K the maximum and minimum temperatures
at the day and night sides, respectively. We set $\tau_{\rm th}$ to
vary linearly with pressure (or $\sigma$) such that, at the $p =
975$~hPa ($\sigma = 0.975$) level, $\tau_{\rm th} = 3.6 \times 10^5$~s
and, at the $p = 25$~hPa ($\sigma = 0.025$) level, $\tau_{\rm th} =
3.6 \times 10^4$~s.  The relaxation time is just slightly longer than
in some recent studies of hot extrasolar planets, making the forcing
slightly less `violent'.  The basic state temperature is isothermal
and set to $T = 1400$~K; and, in all simulations, initial wind {$\bf
  v_0$} is zero everywhere in the computational domain.  The vertical
domain in all the calculations in this test case, as in the previous
test cases, extends from $975$~hPa to 25~hPa (again,
$\approx$~975~mbar to $\approx$~25~mbar, respectively). Similarly,
horizontal resolutions are the same as in TC1 and TC2 and listed with
other model parameters in the Appendix, in
Tables~\ref{table_a7}--\ref{table_a9}.  The calculations are run for
$100 \tau$, much longer than the maximum $\tau_{\rm th}$ ($\approx\!
1.2~\tau$).

To control the small-scale noise inherent in all calculations,
superdissipation (see equation~(\ref{hyper})) is applied in
pseudospectral simulations (in each layer) to prevent accumulation of
energy at the small scales.  Note that this is the `least common
denominator' dissipation, since not all of the tested pseudospectral
cores come with higher order (hyper)viscosity as the default.
Superviscosity is also more equitably compared to explicit viscosity
used in finite volume cores, which in general cannot dissipate as
scale-selectively as cores using the pseudospectral algorithm.  The value of
superdissipation coefficient $\nu_4 = 10^{22}$~m$^4$~s$^{-1}$ at T85
horizontal resolution (corresponding to a damping time of 190~s for
the smallest resolved scale) is chosen based on the study of
\citet{Thrastarson11}: they have found this value to produce a
well-behaved kinetic energy spectrum with the $0.1 \leq (\tau_{\rm
  th}/\tau) \leq 3$ vertical distribution.

In this test case, unlike in TC2 above, the value of
$\nu_{2\mathfrak{p}}$ is increased$\big{/}$decreased with
decreasing$\big{/}$increasing resolution for a given $\mathfrak{p}$ in
pseudospectral calculations (see Table~\ref{table_a7}); this practice
is common in simulation studies.  The procedure will definitely
preclude numerical convergence as we have defined this convergence in
the present work.  However, since our simulations are not numerically
converged in general in the simpler test cases (particularly in TC2),
we do not expect numerical convergence in the more extreme conditions
of TC3.  Hence, we focus our attention on visual and qualitative
convergences in this test case.

In past extrasolar planet simulations performed with MITgcm, it has
been customary to use the Shapiro filter
\citep[e.g.,][]{Showman09,Lewis10}.  Hence, the power-two
($\mathfrak{n} = 2$ in equation~(\ref{shap C})) Shapiro filter is used
here to control oscillations near the grid-scale.  Note that the above
mentioned studies have used $\mathfrak{n} = 4$ Shapiro filter
(Showman, private communication).  Recall that the strength of the
filter is controlled by $\tau_{\rm shap}$, for a given $\Delta t$ and
$\mathfrak{n}$.  In this work, the value of $\tau_{\rm shap}$ is
chosen so that $\Delta t/\tau_{\rm shap} = 1/6$ for all resolutions.
By experimenting with different values, we have found this value to
give flow and temperature structures that are qualitatively in good
agreement with the the pseudospectral cores across different
resolutions: in general, we have found simulations with $\Delta t/
\tau_{\rm shap} = 1/12$ to be under-dissipated and $\Delta t/
\tau_{\rm shap}= 1/3$ to be over-dissipated with the model in its
default setting.

As discussed in Section~\ref{intro}, past simulations of
diabatically-forced, hot, giant extrasolar planets using different
models produce different results---even for fairly similar (but not
identical) setup.  In many cases, the results are qualitatively
different, and the origin of the difference is not obvious.  Quite
often, this is because all the details of the models, model
parameters, and model setup are not reported in the literature---and
sometimes not even described in the original model documentation.
Hence, truly `clean', unambiguous comparisons have not been possible
thus far.  In TC3, the physical setup in all the model calculations is
identical.  Our aim here is to identify whether variation in recent
model results is merely due to the differences in physical setup, or
whether variation is also attributed to differences in the numerical
formulation of a model (e.g., dissipation scheme, spatial grid,
discretization method, etc.).

\subsubsection{TC3 Results}\label{tc3 results}

Fig.~\ref{fig8} shows longitude-latitude maps of the temperature ($T$)
field, with horizontal wind vectors ($\bfv$) overlaid, from
simulations with different model cores.  The resolutions for the
pseudospectral cores and the MITgcm in LL and CS grids are: T85L20,
G128L20, and C64L20, respectively.  The instantaneous fields at the
$p\!  \approx\!  475$~hPa level at $t =
(5\,\tau,\,20\,\tau,\,100\,\tau)$ are shown.  The figure illustrates
the main result of this comparison: when subject to strong
`hot-Jupiter-like' forcing, different model cores produce solutions
which are visually different among them.  This is caused by
spatiotemporal variability in the computed fields and renders specific
predictions, such as the precise location of hot and cold regions,
difficult (see right column of Fig.~\ref{fig8}).

\begin{figure*}
  \vspace*{.5cm}
  \includegraphics[height=20.70cm,width=17.75cm]{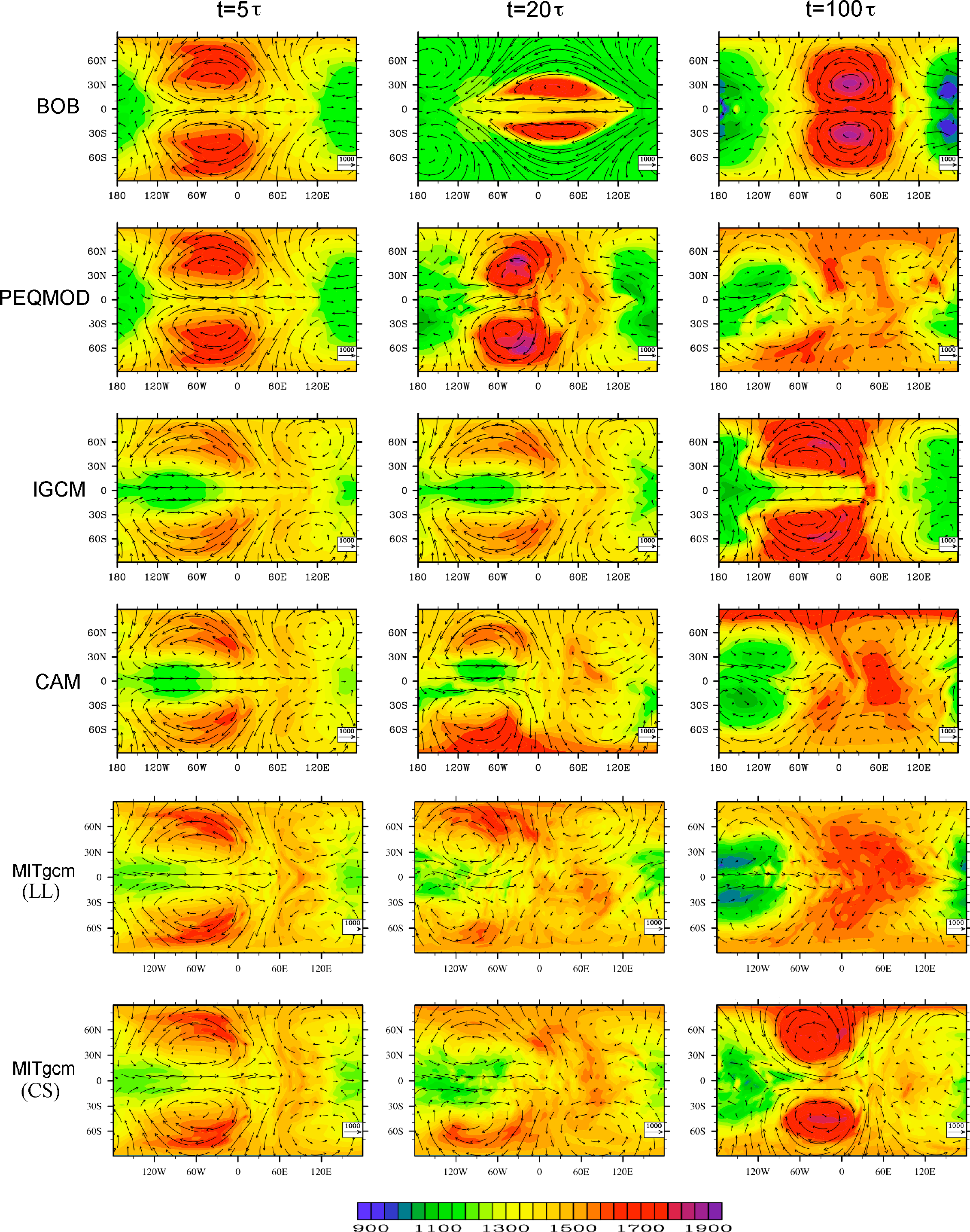}
\vspace*{.1cm}
  \caption{\ Temperature (color coded in K) with wind vectors
    overlaid, for the diabatic forcing test case (TC3) with different
    model cores, at three times at the 475~hPa level.  Form left to
    right, the snapshots are taken at $t = (5\, \tau, 20\, \tau, 100\,
    \tau)$.  The (top, second, third, fourth, fifth, bottom) row is,
    respectively, from a simulation with (BOB, PEQMOD, IGCM, CAM,
    MITgcm in LL grid, MITgcm in CS grid). The resolution for
    pseudospectral codes is T85L20 and for MITgcm in LL and CS grids
    is G128L20 and C64L20, respectively.  The flow and temperature
    distributions are qualitatively similar (e.g., quadrupolar flow)
    but quantitatively different (e.g., time-variable).}
  \label{fig8}
\end{figure*}

Qualitatively, there {\it are} some notable common features.  For
example, most models produce a `quadrupolar-flow' structure, with two
large cyclonic and anti-cyclonic vortex-pairs straddling the equator.
The flow in all the simulations is time variable with vortices
appearing nearly stationary at times or moving longitudinally eastward
or westward at other times, disappearing and reforming on a time scale
of 5--7\,$\tau$.  The temperature in all cases is strongly linked to
the flow and varies on corresponding timescales.  Consequently, the
minimum-to-maximum temperature ranges vary from $\sim$600~K to
$\sim$200~K, at the shown pressure level (e.g., compare middle and
right panel in the third row of Fig.~\ref{fig8}).  Despite the
qualitative similarity, model results are quantitatively very
different and can diverge more markedly when integrated for longer
durations than shown in the figure.

At the beginning, during the first few $\tau$'s, the flow in all
models resembles a linear, `Matsuno-Gill-type' solution
\citep{Matsuno66, Gill80}.  In the solution, westward-propagating
Rossby waves and eastward-propagating Kelvin waves are generated as a
response to the specified mode-1 zonal heating.  At high pressure
(lower altitude) levels, there is a convergent flow
near the substellar point, accompanied by rising motion; concurrently,
there is a divergent flow near the antistellar point, accompanied by
sinking motion (not shown).  At low pressure (higher
  altitude) levels, there is a divergent flow near the substellar
point and a convergent flow near the antistellar point.  In the
classic Matsuno-Gill setup, strong linear (momentum and thermal) drags
balance the forcing.  However, in the absence of strong momentum drag,
as in this test case, nonlinear interactions quickly degrade the
Matsuno-Gill-type solution, and the model solutions start to deviate
strongly from the Matsuno-Gill solution---and, importantly, from each
other.  The latter is due to how different model cores handle
adjustment, as discussed earlier.

At early times ($t < 10\,\tau$), all core solutions are still similar
but a small phase difference is already clearly evident between BOB
and PEQMOD (which are nearly identical to each other at this point)
and the other model cores (see left column in Fig.~\ref{fig8}).  At
later times ($t \ge 10\,\tau$), all solutions start to visibly diverge
from each other and significantly differ quantitatively.  For example,
the north-south flow symmetry is broken in CAM and PEQMOD simulations
(at $t = 14\,\tau$) and in MITgcm in both LL and CS grids (at $t =
10\,\tau$).  In contrast, the symmetry is not broken in BOB and IGCM
cores even at $t = 100\,\tau$.  It is important to understand that the
temporal variability observed here is {\em not} due to large-scale
baroclinic instability stemming from the location and thermal forcing
of the lower boundary.  We have visually and quantitatively checked
(i.e., eddy production, wave propagation, and heat and momentum
fluxes) that large-scale baroclinic instability is not present in this
setup and that similar variability is exhibited even with the lower
boundary placed much deeper (e.g., 10 and 100~bars) with the forcing
limited down to only 1~bar. However, small-scale waves are produced
through the adjustment process.

The flow structure remains either quasi-symmetric about the equator
throughout the integration (see panels for BOB and IGCM in
Fig.~\ref{fig8}) or the equatorial flow symmetry is broken at an early
time (see panels for PEQMOD, CAM, and MITgcm in Fig.~\ref{fig8}).
Detailed analyses of the computed fields show that the symmetry
breaking is associated with emergence of a large equatorial Rossby
wave at $t \approx 10\,\tau$, which is not as prominent in simulations
with BOB or IGCM.  The north-south symmetry in these simulations is
not an artifact of a short integration time. It remains even at the
end of a $2000 \tau$ simulation with BOB at T21L20 resolution (not
shown).  However, when the simulation with BOB is initialized with a
small, random perturbation in the flow, the north-south symmetry does
break at an early time and the flow and temperature evolution closely
match simulations with PEQMOD.  A likely explanation of the equatorial
symmetry breaking is errors introduced by insufficient precision.  By
repeating this test case with PEQMOD at single, double and quadruple
precisions, we have found that the onset time of equatorial symmetry
breaking roughly doubles every time the precision is doubled, in this
experiment.

Since the flow and temperature structure is strongly time variable in
all the simulations, snapshots in time may give an
incomplete---possibly even misleading---picture, since large
differences in Fig.~\ref{fig8} could simply be due to `phasing'
(simple translation of the flow structure in time).  To quantify
variability and the behavior with resolution, a time series of global
average temperature is shown in Fig.~\ref{fig9}.  The first thing to
note is that essentially all the simulations are equilibrated in
temperature: there are no secular growth or decay---hence, the
difference in the fields is not due to failure to achieve `statistical
equilibration'.  The qualitative evolution of global average
temperature is similar in all models after the initial adjustment
period (i.e., for $t \geq 20\,\tau$), with the globally averaged
temperature exhibiting periodic fluctuations of amplitude
$\sim$10--20~K on a timescale of 5--7~$\tau$ (due to the vortex life
cycle discussed above) in all the models.  The exception to this is
the calculations with IGCM: in these calculations, temperature
fluctuates with a clear 10--50\,$\tau$ period, depending on the
resolution, and the amplitude of the fluctuations is much larger
($\sim$\,40~K) than in simulations with other cores.  Note, the
temperature fluctuations in simulations with CAM are also large at
early times ($t \leq 20\,\tau$) but subsequently reduce, as discussed
more in detail below.

\begin{figure*}
  \vspace*{-2.5cm}
  \includegraphics[width=\textwidth]{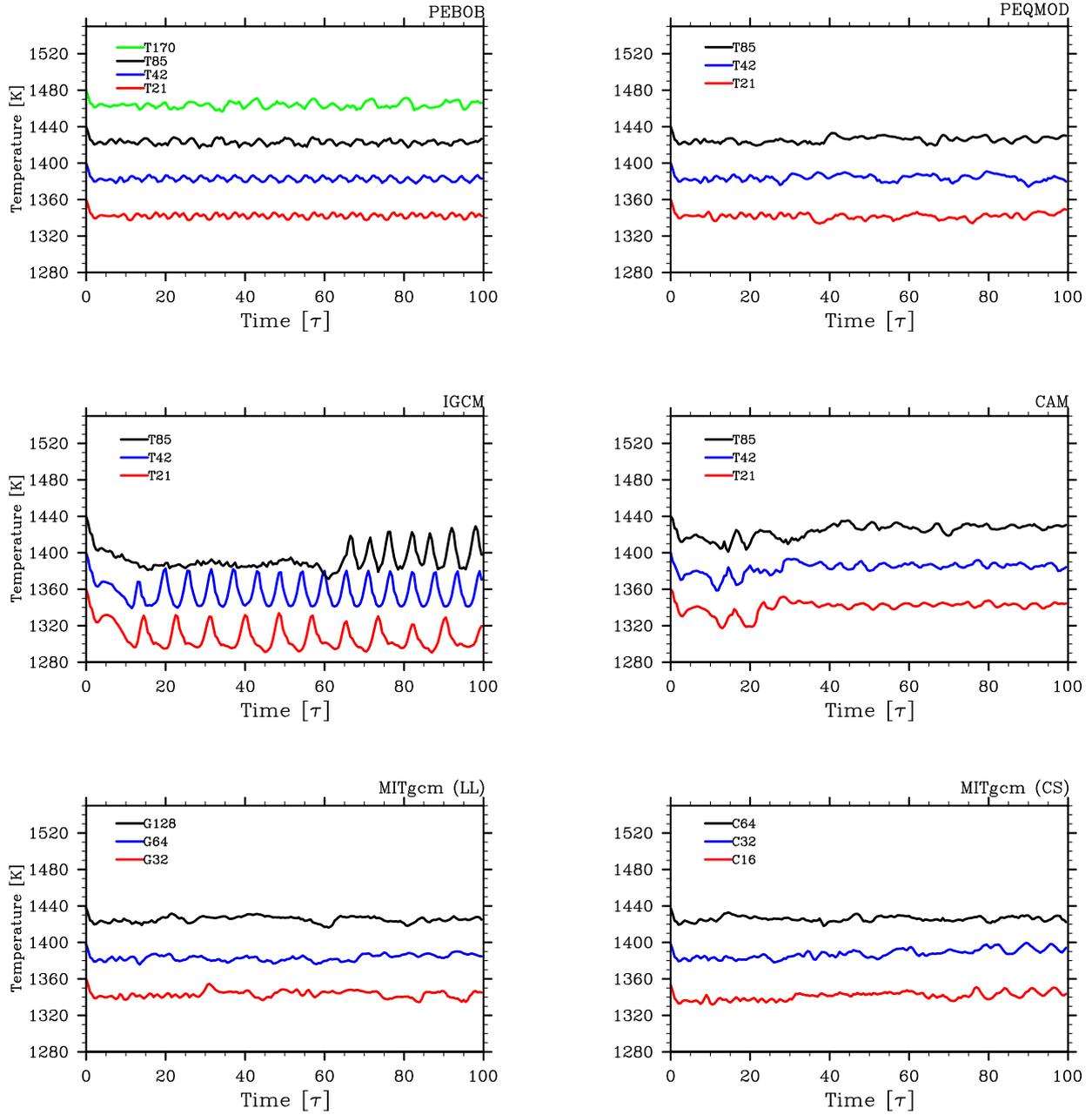}
  \vspace*{-3cm}
  \caption{\ Time series of globally averaged temperature for the
    diabatic test case with different model cores at various
    horizontal resolutions.  The top left panel is from simulations
    with BOB, the top right panel is from simulations with PEQMOD, the
    middle left panel is from simulations with IGCM, the middle right
    panel is from simulations with CAM, the bottom left panel is from
    simulations with MITgcm in LL grid and the bottom right panel is
    from simulations with MITgcm in CS grid.  The curves have been
    offset from each other by 40K, with the temperature of the blue
    color having the correct scale.  The panels show that all the
    calculations are equilibrated in temperature.  They also show that
    qualitatively similar behavior in all the model calculations is
    not due to simple `phasing' of the flow/temperature structures.
    For IGCM, there is a large, non-secular oscillatory behavior. }
  \label{fig9}
  \vspace*{.5cm}
\end{figure*}

Remarkably, the behavior of global average temperature in BOB and
PEQMOD is nearly identical, up to the point when the equatorial
symmetry is broken in PEQMOD (e.g., $t\!\sim\! 30\,\tau$ in T21L20
resolution simulation).  The clear periodicity present in the T21L20
and T42L20 BOB simulations disappears with increasing horizontal
resolution (cf.~red curve with green curve in top left panel of
Fig.~\ref{fig9}, for example).  Note that, in all simulations, the
global average temperature decreases by 10--20~K from the initial
value of 1400K.  The initial dip is related to the short timescale on
which the large $T_e$ gradient is relaxed. For example, if the
relaxation time is increased by 1~$\tau$ everywhere in a BOB core
simulation, the global average temperature decreases
by less than 1~K (not shown). The adjustment demanded
by the fast relaxation produces violent flows.  Such a representation
of thermal forcing is not physical---certainly its resulting flow is
difficult to model accurately in current GCMs.  Nevertheless, since we
are primarily concerned with comparing the model cores, we use the
representation here---for heuristic purposes.
 
Large amplitude fluctuations in the global average temperatures of
IGCM and CAM simulations (particularly at early times in the latter)
are associated with atmospheric thickness variations, caused by
fluctuations in the surface pressure (recall that both cores use the
$\sigma$-coordinate).  This `flapping' of the bottom boundary is
absent in $p$-coordinate models with rigid top and bottom boundaries,
in which surface pressure remains constant throughout the integration.
By removing the bottom boundary away from the forcing region, we have
found the atmospheric thickness variation to be greatly reduced in
$\sigma$-coordinate models.  Interestingly, as already mentioned, the
fluctuations in the CAM core subside at $t \gtrsim 30\,\tau$.  This is
likely due to the $\eta$-coordinate employed by the core.

Note that the flapping is not necessarily unphysical, and could be
used to represent a physical phenomena at the 1~bar level if a hot
extrasolar planet happens to have a natural, non-rigid boundary (e.g.,
jump in stratification, composition etc.)  there.\footnote{Such a
    setup is common in Earth's middle atmosphere and climate studies
    to represent thermal and mechanical forcing.} It is also possible
to specify a free-surface boundary condition at the bottom in MITgcm.
The specification replaces the condition, $\omega = 0$ at $p = p_{\rm
  r}$, with $\omega = \D p_s/\D t$ at $p = p_{\rm r}$.  With this
boundary condition an additional prognostic equation for free surface
pressure anomaly, called $\hat{\eta}$ in MITgcm (not to be confused
with the $\eta$ associated with the CAM), is solved.  However, given
the strong forcing and violent flow that ensues, MITgcm in the default
setting crashes with the boundary at 1~bar with the free-surface
condition.  This is due to the large undulation of the pressure
surface, which can cause two or more pressure surfaces to intersect
somewhere in the domain.  Loss of single-valuedness such as this is
not an issue in $\sigma$-coordinate model cores because surface
pressure is not a coordinate surface.

It is clear from Fig.~\ref{fig9} that the cores are not numerically or
visually converged. This is not surprising, especially for the
pseudospectral cores, given that the superdissipation coefficient is
decreased with increasing horizontal resolution.  However, qualitative
convergence in TC3 appears to be achieved at the lowest horizontal
resolutions (i.e., T21, G32 and C16) in all the cores: at least, the
qualitative behavior of flow and temperature appears to be the same at
all resolutions.  Given the behavior observed in TC2, however, we
cannot completely rule out that this conclusion may need to be revised
when simulations of substantially higher resolution (than those in
this work) are carried out. In any case, we note here that the
conclusion depends on the field or quantity considered, as we shall
show later.  For example, qualitative convergence is not achieved when
the vorticity field is considered instead \citep{Thrastarson11}.

In TC3, the total global energy is not conserved because $\dot{q}/c_p
\ne 0$ in the thermodynamic equation.  However, in the absence of
momentum forcing, the global integral of absolute
angular momentum is still conserved.  While the angular momentum
conservation has been almost exact in TC1 and TC2 (up to
$\sim$\,0.02~percent), in this test case the conservation property is
strongly violated in some model cores---particularly in their default
configurations.  This is shown in Fig.~\ref{fig10}, which presents
time series of globally integrated absolute angular momentum
(equation~(\ref{AM})).  Each time series is normalized by its initial
value of $1.77\!  \times\!  10^{32}$~kg~m$^2$~s$^{-1}$.

\begin{figure*}
  \vspace*{-2.5cm}
  \includegraphics[width=\textwidth]{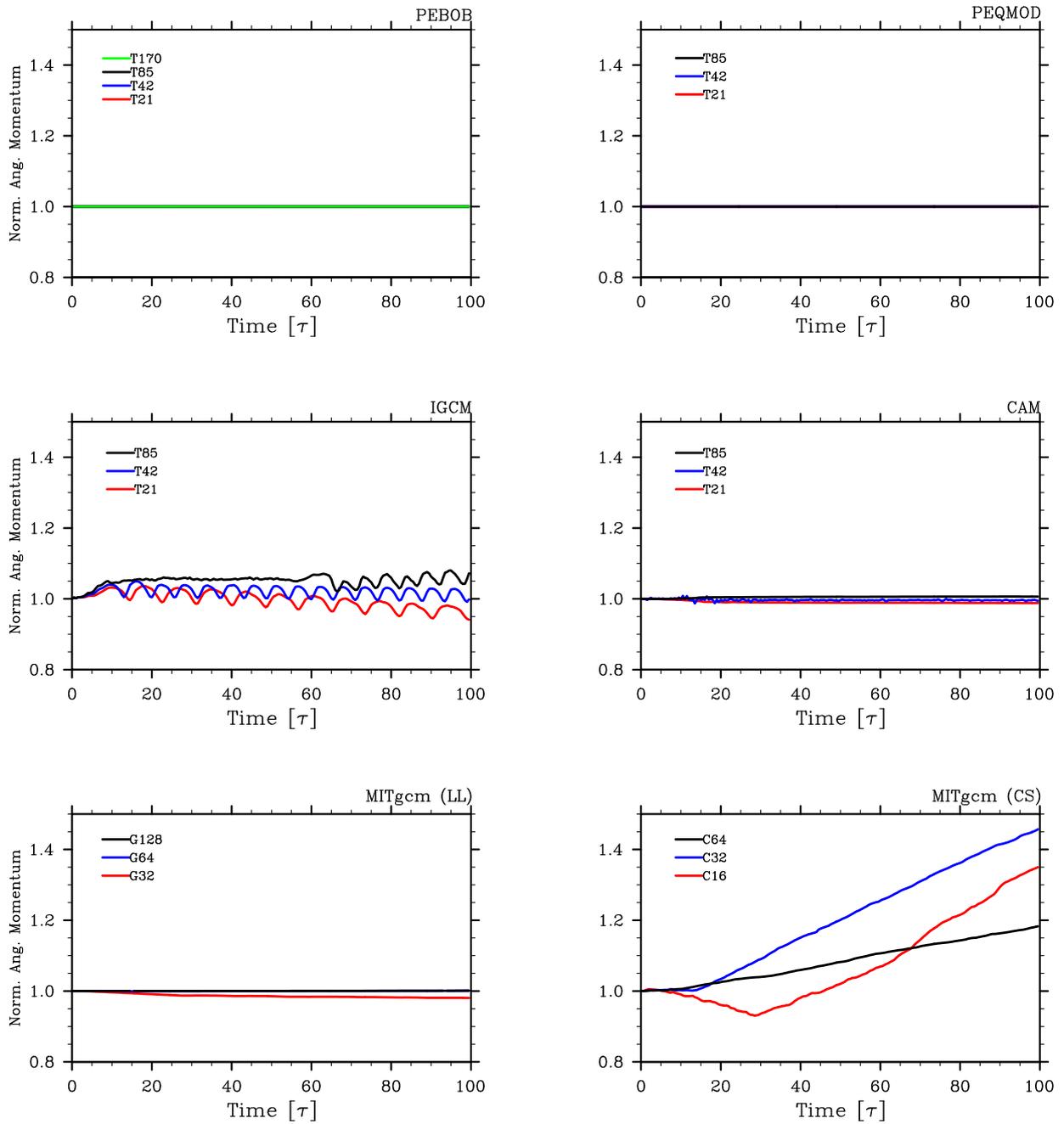}
  \vspace*{-2.75cm}
  \caption{\ Time series of global angular momentum (normalized by
    initial value of 1.77$\times 10^{32}$~kg~m$^2$~s$^{-1}$) for the
    diabatic test case with different model cores at various
    horizontal resolutions.  The panel placements are as in
    Fig.~\ref{fig9} (i.e., BOB core at upper left, etc.).  The total
    absolute angular momentum is exactly conserved at all resolutions
    in PEBOB and PEQMOD and at high resolutions in MITgcm in LL grid.
    It is somewhat poorly conserved in IGCM and not at all in MITgcm
    in CS grid. }
  \label{fig10}
  \vspace*{.2cm}
\end{figure*}

As can be seen from Fig.~\ref{fig10}, only BOB and PEQMOD conserve
angular momentum exactly---{\it and they do so at all resolutions}
(see top row of the figure).  The conservation property of CAM and
MITgcm in LL grid becomes better with increasing resolution.  The
periodic `spinning' and 'de-spinning' of the atmosphere in IGCM by
$\sim$10~percent (see middle left panel of Fig.~\ref{fig10}) is
caused by variation in the atmospheric mass, due to large surface
pressure variations.  Clearly, angular momentum is poorly conserved in
the calculations with the MITgcm in the CS grid,
especially at C32 resolution: the total angular momentum increases by
more than 45~percent at the end of the $100\,\tau$ simulation. The
conservation is actually better at the lower, C16 resolution, in terms
of the time series at $t = 100\,\tau$.  When these simulations are
integrated for longer than $100\,\tau$, the runaway angular momentum
is associated with a zonally-averaged zonal flow which is {\it
  strongly} superrotating\footnote{i.e., $\overline{u}\, >\,
  (\Omega\,R_{\rm p}\, \sin^2\phi/ \cos \phi)$, where $\overline{u} =
  \overline{u}(\phi,s,t)$ is the zonally-averaged wind.  Here,
  `strongly' means $\overline{u}$ close to, or even exceeding,
  3000~m~s$^{-1}$.} over a broad range of latitudes.  For example, the
transition to such a superrotating state occurs by $t \approx
200\,\tau$ in C16 and C32 simulations.

The angular momentum runaway behavior in the MITgcm in CS grid
simulations is caused by an instability associated with the Shapiro
filter used in the simulations.  When the strength of the filter is
doubled (i.e., $\Delta t/\tau_{\rm shap} = 1/3$), angular momentum
decreases over time leading to unphysical {\it subrotation} (i.e.,
westward flow of the atmosphere everywhere).  This occurs by $t =
100\,\tau$.  When the strength is halved (i.e., $\Delta t/\tau_{\rm
  shap} = 1/12$), not enough dissipation is supplied to the flow and
the runaway still occurs and is more severe at an earlier time.
Indeed, through an extensive study, we have found that a suitable
strength of the Shapiro filter (which would conserve angular momentum
exactly on the cubed sphere) does not exist for TC3.  Note that this
behavior is not just limited to the $\mathfrak{n} = 2$ Shapiro filter.
It occurs for higher power filters (e.g., $\mathfrak{n} = 4$ and
$\mathfrak{n} = 6$) as well.  Note also that, for this test case, the
MITgcm in CS grid at C16 resolution (in the default setting) always
crashes with $\mathfrak{n} = 8$ Shapiro filter, independent of $\Delta
t/\tau_{\rm shap}$.

For completeness, we have also performed TC3 with the MITgcm core in
both LL and CS grids (at G32 and C16 resolutions, respectively) with
ordinary, Laplacian dissipation ($\nabla^2$).  The
damping time in these simulations is chosen to be the same as in the
simulations performed with the pseudospectral cores.  The use of
ordinary dissipation considerably improves the angular momentum
conservation of MITgcm in CS grid (a monotonic increase of only
$\sim$0.5~percent at $t = 100\,\tau$).  However, there is now a
significant ($\sim$40~percent) loss of angular momentum in the LL
grid simulation using the same dissipation.  In other words,
the CS grid simulation is severely overdissipated
compared to the LL grid simulation.  From this we conclude that the
dissipation is compensating the runaway in the CS grid.  Therefore,
the grid itself appears to be a significant source of the runaway
behavior, with the Shapiro filter amplifying the behavior---at least
in the model's default CS grid configuration.}

\begin{figure*}
  \vspace*{-2.75cm}
  \includegraphics[width=\textwidth]{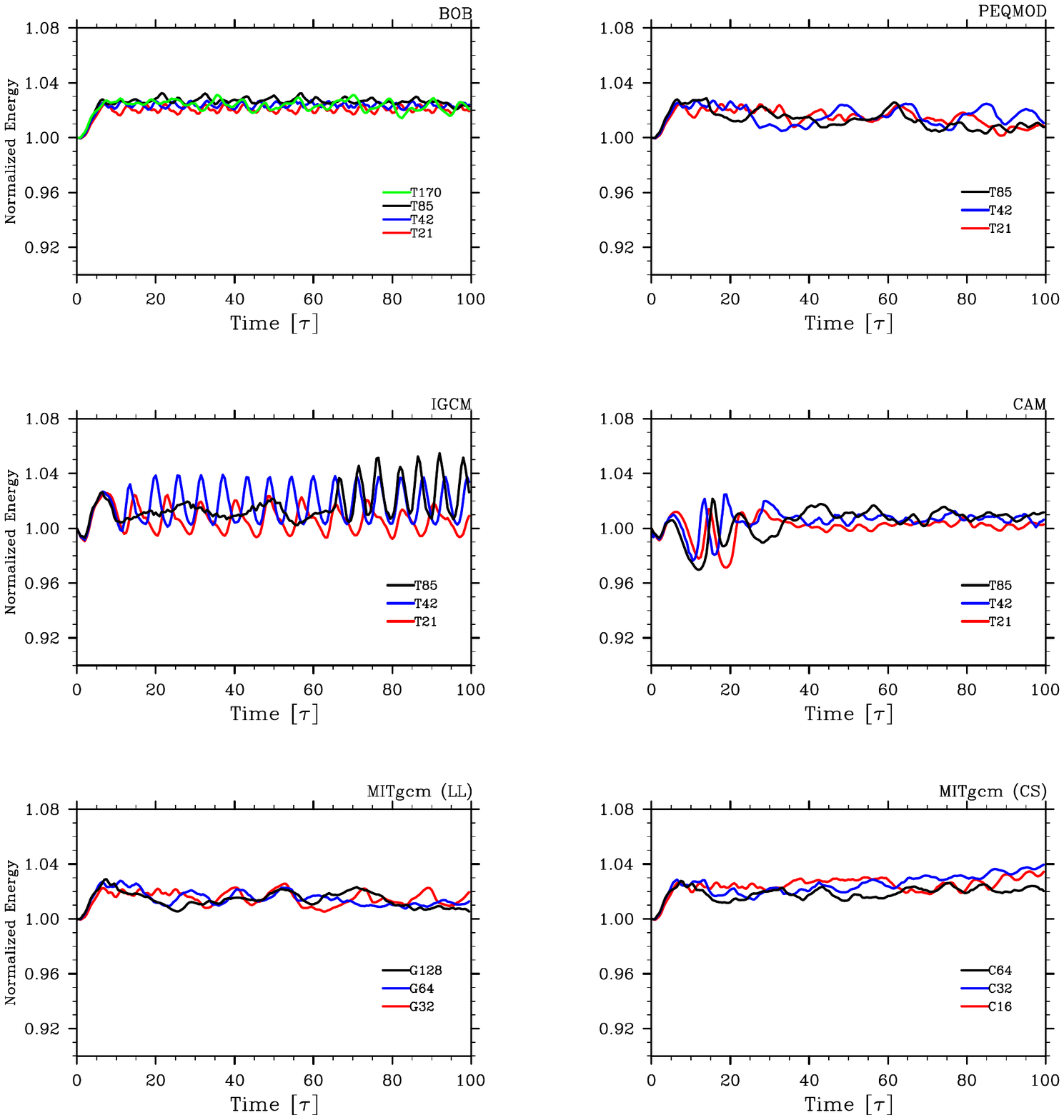}
  \vspace*{-3cm}
  \caption{\ Time series of global total energy (normalized by initial
    value of 2.2$\times 10^{28}$~J) for test case~3 with different
    model cores at varying horizontal resolutions.  The panel
    placements are as in Fig.~\ref{fig9} (i.e. BOB core at upper left,
    etc.).  All the simulations appear to be well or roughly
    equilibrated, with the possible exception of the lower resolution
    simulations with the MITgcm core in CS grid.  The total energy
    fluctuations are correlated with the temperature fluctuations
    (cf.~Fig.~\ref{fig9}). }
  \label{fig11}
  \vspace*{.2cm}
\end{figure*}

As already discussed, the total energy with the applied forcing in TC3
is not expected to be conserved by any of the models.  However, as
shown below, the inclusion of Newtonian relaxation does not alter the
total atmospheric energy budget by more than 5 percent.
Fig.~\ref{fig11} shows a global integral of total energy as a function
of time, normalized by the initial value of 2.2$\times 10^{28}$~J. In
all but the IGCM and CAM simulations the total energy steadily
increases at the beginning by $\sim$2.5 percent.  Thereafter, BOB
maintains the total energy at a nearly constant level (see top left
panel of the figure).  With the other cores, the total energy
fluctuates by up to 5~percent, with largest fluctuations observed in
IGCM simulations; at $t = 100\,\tau$ the total energy for IGCM
increases by 2--3~percent, on the average.  The energy in the MITgcm
simulations in CS grid also increases noticeably over time
(particularly in the lower resolution simulations), consistent with
the runaway angular momentum and transition to a superrotating state
for $t \gtrsim 200\,\tau$, discussed above.  The total energy for CAM
shows a slight increases, by $\sim$1~percent, over the initial value,
after settling from large initial fluctuations; but, PEQMOD and MITgcm
LL grid simulations show the energy decrease slightly by $\sim$1.5
percent, after the initial rise of $\sim$2.5~percent mentioned above.
 
In the process of thoroughly verifying our results, we have
discovered, as already noted, two additional advection schemes for the
vector invariant momentum equation for the MITgcm CS grid.  Of the two
undocumented schemes, one explicitly conserves energy
\citep{Sadourny75} and the other conserves energy {\it and} enstrophy
\citep{Burridge77}.  While detailed discussion on the behaviours of
MITgcm with these two `non-default' schemes are reported elsewhere
(Polichtchouk and Cho, in prep.), we note that the angular momentum
conservation is noticeably improved in TC3, particularly with the
energy conserving scheme. For example, with the use of the energy
conserving scheme the angular momentum in TC3 {\it decreases} by
33~percent at C16 resolution, by 6~percent at C32 resolution and by
2~percent at C64 resolution at the end of the $100\,\tau$
simulation. Because these decreases are monotonic with time longer
time integrations would result in a more significant violation of
angular momentum consevation in TC3. We also note that the two
undocumented schemes do not improve the performance in TC1 and TC2,
compared to the default enstrophy conserving scheme.

As a closing remark, somewhat disconcertingly, we have noticed that
the model cores are perceptibly sensitive to small, perhaps
`uncontrollable', changes in input parameters when subject to the
strong forcing, as in TC3.  We have already shown that the simulation
results are different 1) between different model cores at the same or
comparable resolution and 2) between different resolutions with the
same model core.  However, the simulation results can also differ with
the same core at the same resolution.  For example, simulations with
PEQMOD at T42L20 resolution with single, double and quadruple
precision produce quantitatively unconverged results (although they
may be qualitatively converged).  In addition to the onset of the
north-south symmetry breaking, we have found a noticeable phase
difference between the solutions.  Such phase differences are
significant for predicting precise flow and temperature patterns.
Moreover, we have found the phase differences to occur in a model core
with identical setup and precision when the core was simply compiled
with a different compiler or the same compiler on a different
computing platform.  It is important to note that the above issues did
not play significant roles in TC1 and TC2 because `violent' (i.e.,
strongly unbalanced) flows are not involved.  Given this, we do not
expect the results presented in TC3 to be {\it exactly} reproducible
when a different compiler is used and/or on different platforms:
slight phase variations are nominally expected.

\section{Summary and Discussion}\label{summary}

Intercomparison and benchmarking of GCMs, and in particular their
cores, are necessary for assessing the efficacy of models and for
understanding the physical properties of atmospheres.  While such
testing is common practice in Earth and some Solar System planet
studies, only three tests \citep[e.g.,][]{Rauscher10, Heng11,
  Bending13} have been attempted for hot extrasolar planet GCMs.
Here, we subject five GCMs currently used in the hot extrasolar planet
studies to three benchmark tests.  The tested GCMs are: BOB, PEQMOD,
IGCM, CAM, and MITgcm.  These models employ a range of numerical
algorithms for the spatial discretization and explicit viscosity:
respectively, pseudospectral with pseudospectral filtering (BOB,
PEQMOD, IGCM, and CAM) and finite volume with differenced, pole
(zonal), and Shapiro filtering (MITgcm).  All the GCMs solve the dry,
hydrostatic primitive equations and are subjected to identical tests.
From least to most stringent, these tests are: 1) the steady-state
test case (TC1); 2) the baroclinic wave test case (TC2); and, 3) the
diabatic test case (TC3).  In all three test cases, all of the models
are tested at varying horizontal resolutions to assess numerical
convergence.  Both inter- and intra-comparisons are carried out.

TC1 assesses how a model is able to maintain a balanced initial
condition in the form of a midlatitude eastward jet with no applied
dissipation, before gravity waves and numerical noise degrade the jet.
With the exception of MITgcm employing the CS grid, all models
maintain the true, steady-state solution very well throughout the time
of integration ($t = 20 \, \tau$).  The special corners in the CS mesh
quickly degrade the balanced state; and, in the absence of any
dissipation, the imbalance causes the model simulations to crash at
early time when using the default advection scheme{\footnote{Note that
    the steady-state simulations with energy and energy {\it and}
    enstrophy conserving schemes do not crash before $t = 20 \,
    \tau$. However, the physical space fields are inundated with
    grid-scale noise at early time and the simulations are clearly
    unbalanced.}: for example, the simulation at C64 resolution
  crashes at $t = 4.5\, \tau$.  Of all the models, BOB and PEQMOD
  maintain the true solution the best (see, e.g., Fig.~\ref{fig3}).
  
In TC2, a temperature perturbation applied to the steady-state initial
condition of TC1 triggers a nonlinear evolution of a baroclinic wave.
The emergence of sharp fronts from baroclinic instability necessitates
the use of flow viscosity, unlike in TC1.  We have chosen ordinary
Laplacian (i.e., $\nabla^2$) dissipation.  While all model
calculations permit baroclinic instability, only pseudospectral ones
appear to be visually converged, and this occurs at a horizontal
resolution of T85 (corresponding to 85 total and zonal modes each in
the spherical harmonic expansion---a resolution above most current
extrasolar planet simulations).  Solutions with MITgcm are not
converged at the highest resolutions investigated here (i.e., G128 and
C64).  While the most unstable mode of the unstable jet is
approximately 3--4 (see Fig.~\ref{fig4} and Fig.~\ref{fig5}), the
presence of the special corner points on the cubed-sphere grid
produces an unstable wave field with mode greater than 3--4 (see
Fig.~\ref{fig5} bottom right panel).  This behavior is significant for
model predictions and observational confirmations.

While the first two test cases are adiabatic, the third case is
diabatic.  For some regions of the planet atmosphere, diabatic forcing
is necessary.  In this work, as in most hot extrasolar planet studies
thus far, we apply a simple Newtonian relaxation parameterization to
represent the heating and cooling in the modeled atmosphere.  Here,
the atmosphere is `spun-up' from an initial isothermal condition at
rest.  Note that, although many past studies have employed a similar
setup, there is no general agreement on the robustness of the flow and
temperature distributions produced by the simulations.  By employing
an identical setup in all the models, the aim of TC3 has been to
clearly assess whether (and how much) the non-robustness of the
results is intrinsic to the numerical model employed.  We emphasize
again that without such tests, the simulation community---indeed the
extrasolar planet community as a whole---would not have a baseline for
any consensus.  For this reason, we employ a biharmonic ($\nabla^4$)
superviscosity in pseudospectral models and a power-two ($\mathfrak{n}
= 2$) Shapiro filter in MITgcm so that our comparisons are equatable
among the models tested in this study and shed some light on the
results of past studies. \footnote{We remind the reader here that the
  MITgcm in this default CS grid configuration possesses the least
  amount of angular momentum runaway at $t = 100\,\tau$ with
  $\mathfrak{n} = 2$.}

Unlike in the first two test cases, the extreme forcing condition of
the third test case produces a range of behaviors in the model
calculations.  While there are some qualitatively similar features
(e.g., a time-variable quadrupole flow structure), the location and
magnitude of the hottest and coldest regions are not same in the model
calculations.  In large part, this is due to significant phase
differences in the computed fields.  Moreover, apart from BOB and
IGCM, all models break the flow symmetry about the equator relatively
early on, before $t = 100\, \tau$ (see Fig.~\ref{fig8}).  This
behavior may be somewhat surprising as the forcing is north-south
symmetric and no initial noise is present to break the symmetry, but
machine precision or coding inexactitude eventually break the symmetry
in all the cores tested.

Throughout this work we have paid careful attention to the
conservation properties of the numerical models.  In TC1 and TC2, the
angular momentum and total energy conservations are well fulfilled by
all the models.  In TC3, however, we have found that---apart from BOB
and PEQMOD---no other model conserves angular momentum exactly.  The
MITgcm in CS grid with Shapiro filter perform the poorest in this
case, with the total absolute angular momentum increasing by as much
as 45 percent at the end of the integration ($t =100\, \tau$).  As
pointed out in \citet{Thrastarson11} and \citet{Polichtchouk12}, the
normally `relatively harmless' small non-conservation in Earth-like
conditions, for which the GCMs have been constructed and tested, could
be exacerbated in the hot extrasolar planet condition.  This is
because of the exacting requirements the condition places on the
numerics.  Such model behaviors should be carefully taken into
account, when performing hot extrasolar planet simulations.  We
emphasize this point in this work because we note that runaway (or
decaying) angular momentum is not necessarily apparent from looking at
the flow pattern alone (see Fig.~\ref{fig8} bottom row).  Long-time
integration of such a simulation leads to an atmosphere that
superrotates in a manner similar to those reported in previous
studies.

GCMs are complex.  Getting them to run properly and verifying their
results is not trivial.  Trade-offs between accuracy, speed and
algorithmic/coding simplicity are always made; and, even when such
things are well-understood theoretically, the actual behavior of the
model is not always `stable' or uniform across problems.  This then
also raises the complexity of interpreting the results.  In this work,
we have endeavoured to fairly assess the performance of a number of
GCMs---covering a good cross-section of algorithms, grids and
treatment of explicit viscosity.  For the most part, the tested models
behave well and similarly to each other, but with some unexpected
results.  Although we have extensively checked our calculations here
and despite our best effort to provide all the details of the
calculations, it is still possible---especially given some of the
findings presented in TC3---that other studies may obtain different
results.  Broadly, we offer this work for other studies to reproduce
and to build on, as it seems clear to us that that is necessary for
advancement in theory and modelling of hot extrasolar planet
atmospheres.

\section*{Acknowledgments}

The authors thank three anonymous reviewers for helpful comments,
which improved this manuscript. This work has been supported by the
Science and Technology Facilities Council (STFC) research studentships
to I.P. and C.W., STFC grant PP/E001858/1 and Westfield Small Grant
(WSG) to J.Y-K.C.  H.T.T was supported by an appointment to the NASA
Postdoctoral Program at the Jet Propulsion Laboratory, administered by
Oak Ridge Associated Universities through a contract with
NASA. H.T.T. was also supported in part by the National Science
Foundation under Grant No. PHYS-1066293 and the hospitality of the
Aspen Center for Physics. O.M.U. and M.T.J. acknowledge support from
STFC and WSG, respectively.

\appendix

\section{Tables of Values for Test Cases}

\setcounter{table}{0}
\begin{table*}[ht]
\centering
  \caption{Table of vertical and horizontal grid resolutions as well 
    as other parameters needed for reproduction of steady-state case 
    with pseudospectral cores.  Note, PEQMOD has an additional 
    equatorial latitude point (i.e. T21 pseudospectral resolution 
    corresponds to $33 \times 64$ grid points, T42 to $65 \times 128$ 
    grid points etc.). BOB only has been integrated at resolution 
    T170L20.}
  \label{table_a1}
 \scalebox{1.0}{
  \begin{tabular}{ccccccc} 
    \\
    \hline
    Horizontal & Vertical & Gaussian Grid & Timestep & Hyperdissipation  
    & Robert-Asselin \\
    Resolution & Resolution & (lon $\times$ lat) & ($\Delta t$)\ \ [s]
    & & Coefficient\ \ ($\epsilon$)  \\
    \hline
    T21  & L20   & $64 \times 32$    & 120  &  No  &  0.001 \\
    T42  & L20   & $128 \times 64$   & 60   &  No  &  0.001 \\
    T85  & L20   &  $256 \times 128$ & 30   &  No  &  0.001 \\
    T170 & L20   &  $512 \times 256$ & 15   &  No  &  0.001 \\ 
    \hline \\*[.25cm] 
  \end{tabular}}
\end{table*}

\begin{table*} 
  \centering
  \caption{As in Table~\ref{table_a1}, but for MITgcm in 
    longitude-latitude (LL) grid.  Note, the computational grid 
    size below should not be compared directly with the Gaussian 
    grid size in Table~\ref{table_a1}, as the 
    pseudospectral evolves the fields in spectral space: the Gaussian 
    grid is used only to evaluate the nonlinear products and de-alias 
    the fields.}
  \label{table_a2}
\scalebox{1.0}{
  \begin{tabular}{ccccccc} 
    \\
    \hline
    Horizontal & Vertical & Computational Grid & Timestep 
    & Harmonic & Zonal (FFT) & Shapiro \\
    Resolution & Resolution & (lon $\times$ lat) & ($\Delta t$)\ \ [s]
    & Dissipation & Filter& Filter\\
    \hline
    G32   & L20  & $64 \times 32$ &120  &No& Yes &No\\
    G64   & L20  & $128 \times 64$ & 60 & No& Yes &No \\
    G128  & L20  &  $256 \times 128$ &30 &No& Yes &No\\
    \hline\\ 
  \end{tabular}}
\end{table*}

\begin{table*} 
  \centering
  \caption{Same as Table~\ref{table_a1}, but for MITgcm in
    cubed-sphere (CS) grid.  N.B., the C16 and C64 CS grids have been
    generated by us with MATLAB routines provided by MITgcm support;
    but, the MATLAB generated C32 grid comes included with the
    MITgcm.}
  \label{table_a3}
 \scalebox{1.0}{
  \begin{tabular}{ccccccc} 
    \\ \hline Horizontal & Vertical & Computational Grid & Timestep &
    Harmonic & Zonal (FFT) & Shapiro \\ Resolution & Resolution &
    (irregular) & ($\Delta t$)\ \ [s] & Dissipation & Filter & Filter
    \\ \hline C16 & L20 & $6 \times 16 \times 16$ & 120 & No & No & No
    \\ C32 & L20 & $6 \times 32 \times 32$ & 60 & No & No & No \\ C64
    & L20 & $6 \times 64 \times 64$ & 30 & No & No & No
    \\ \hline\\*[.25cm]
  \end{tabular}}
\end{table*}

\begin{table*}
  \centering
  \caption{Table of vertical and horizontal grid resolutions as well 
    as other parameters needed for reproduction of baroclinic wave 
    test case with pseudospectral cores. Note, as above, PEQMOD has 
    an additional equatorial latitude point (i.e. T21 pseudospectral 
    resolution corresponds to $33 \times 64$ grid points, T42 to 
    $65 \times 128$ grid points etc.). Also, note, only BOB has been 
    integrated at resolution T170L20.}
  \label{table_a4}
\scalebox{1.0}{
  \begin{tabular}{cccccccc} 
    \\
    \hline
    Horizontal & Vertical  & Guassian Grid & Timestep 
    & Hyperdissipation & Dissipation & Dissipation Coefficient 
    & Robert-Asselin\\
    Resolution & Resolution & (lon $\times$ lat) & ($\Delta t$)\ \ [s] 
    & & Order\ \ ($\mathfrak{p}$) & $(\nu_2)$\ \ [m$^2$~s$^{-1}$]  
    & Coefficient\ \ ($\epsilon$) \\
    \hline
    T21  & L20  & $64 \times 32$   & 120 & Yes & 1 & $2\times10^7$ & 0.001\\
    T42  & L20  & $128 \times 64$  & 60  & Yes & 1 & $2\times10^7$ & 0.001\\
    T85  & L20  & $256 \times 128$ & 30  & Yes & 1 & $2\times10^7$ & 0.001\\
    T170 & L20  & $512 \times 256$ & 15  & Yes & 1 & $2\times10^7$ & 0.001\\ 
    \hline\\*[.25cm] 
  \end{tabular}}
\end{table*}

\begin{table*} 
  \centering
  \caption{Same as Table~\ref{table_a4}, but for MITgcm in LL grid.}  
  \label{table_a5}
 \scalebox{1.0}{
  \begin{tabular}{ccccccccc} 
    \\ \hline Horizontal & Vertical & Computational Grid & Timestep &
    Harmonic & Dissipation Coefficient & Zonal (FFT) & Shapiro
    \\ Resolution & Resolution & (lon $\times$ lat) & ($\Delta
    t$)\ \ [s] & Dissipation & ($\nu_2$) [m$^2$~s$^{-1}$] & Filter &
    Filter\\ \hline G32 & L20 & $64 \times 32$ &120 & Yes
    &$2\times10^7$ & Yes &No\\ G64 & L20 & $128 \times 64$ &60 & Yes
    &$2\times10^7$ & Yes &No\\ G128 & L20 & $256 \times 128$ &30 & Yes
    &$2\times10^7$ & Yes &No\\ \hline\\*[.25cm]
  \end{tabular}}
\end{table*}

\begin{table*} 
  \centering
  \caption{Same as Table~\ref{table_a4} but for
    MITgcm in CS grid.}
  \label{table_a6}
\scalebox{1.0}{
  \begin{tabular}{cccccccccc} 
    \\ \hline Horizontal &Vertical & Computational Grid & Timestep &
    Harmonic & Dissipation Coefficient & Zonal (FFT) &
    Shapiro\\ Resolution & Resolution & (irregular) & ($\Delta
    t$)\ \ [s] & Dissipation & ($\nu_2$) [m$^2$~s$^{-1}$] &Filter &
    Filter \\ \hline C16 &L20 & $6\times 16 \times 16$ &120& Yes&
    $2\times 10^7$&No &No \\ C32 &L20 & $6\times 32 \times 32$ & 60 &
    Yes& $2\times 10^7$&No &No \\ C64 &L20 & $6\times 64 \times 64$
    &30 & Yes& $2\times 10^7$&No &No \\ \hline\\*[.25cm]
  \end{tabular}}
\end{table*}

\begin{table*}
  \centering
  \caption{Table of vertical and horizontal grid resolutions as well 
    as other parameters needed for reproduction of diabatic forcing
    test case with pseudospectral cores. Note, as above, PEQMOD has 
    an additional equatorial latitude point (i.e. T21 pseudospectral 
    resolution corresponds to $33 \times 64$ grid points, T42 to 
    $65 \times 128$ grid points etc.). Here also, only BOB has been 
    integrated at resolution T170L20.}
  \label{table_a7}
\scalebox{1.0}{
  \begin{tabular}{cccccccccc} 
    \\
    \hline
    Horizontal & Vertical & Gaussian Grid & Timestep & Hyperdissipation &
    Dissipation& Dissipation Coefficient &    Robert-Asselin\\
    Resolution & Resolution & (lon $\times$ lat) & ($\Delta t$) \ \ [s]&
    & Order ($\mathfrak{p}$) & $\nu_4$~[m$^4$~s$^{-1}$] & 
    Coefficient ($\epsilon$)\\
    \hline
    T21 & L20  & $64 \times 32$ &240  &Yes&2 & $1\times10^{23}$& 0.01\\
    T42 & L20  & $128 \times 64$ & 120 & Yes&2 & $5\times10^{22}$&0.01 \\
    T85 & L20  &  $256 \times 128$ &60 & Yes&2& $1\times10^{22}$&0.01\\
    T170& L20  &  $512 \times 256$ &30  & Yes&2& $5\times10^{21}$&0.01\\ 
    \hline\\*[.25cm] 
  \end{tabular}}
\end{table*}

\begin{table*} 
  \centering
  \caption{Same as Table~\ref{table_a7} but
    for MITgcm in
    LL grid.}
  \label{table_a8}
\scalebox{1.0}{
  \begin{tabular}{ccccccccc} 
    \\ \hline Horizontal & Vertical & Computational Grid & Timestep &
    Harmonic & Zonal (FFT) & Shapiro & Filter & Filter Parameter
    \\ Resolution & Resolution & (lon $\times$ lat) & ($\Delta
    t$)\ \ [s] & Dissipation & Filter & Filter & Power
    ($\mathfrak{n}$) & ($\tau_{\rm shap}$) \ \ [s]\\ \hline G32 & L20
    & $64 \times 32$ &240 & No & Yes &Yes & 2& 1440 \\ G64 & L20 &
    $128 \times 64$ &120 & No & Yes &Yes & 2& 720\\ G128 & L20 & $256
    \times 128$ &60 & No & Yes &Yes & 2& 360\\ \hline\\*[.25cm]
  \end{tabular}}
\end{table*}
\begin{table*} 
  \centering
  \caption{Same as Table~\ref{table_a7} but for MITgcm in CS grid.}
    \label{table_a9}
\scalebox{1.0}{
   \begin{tabular}{ccccccccc} 
     \\ \hline Horizontal & Vertical & Computational Grid & Timestep &
     Harmonic & Zonal (FFT) & Shapiro & Filter & Filter Parameter
     \\ Resolution & Resolution & (irregular) & ($\Delta t$)\ \ [s] &
     Dissipation & Filter & Filter & Power ($\mathfrak{n}$) &
     ($\tau_{\rm shap}$) \ \ [s]\\ \hline C16 & L20 & $6\times 16
     \times 16$ &240 & No & No &Yes & 2& 1440 \\ C32 & L20 & $6\times
     32 \times 32$ &120 & No & No &Yes & 2& 720\\ C64 & L20 & $6\times
     64 \times 64$ &60 & No & No &Yes & 2& 360\\ \hline\\*[.25cm]
    \end{tabular}}
\end{table*}

\end{document}